\documentclass[fleqn,10pt]{SelfArx} % Document font size and equations flushed left

\definecolor{color1}{RGB}{0,0,90} % Color of the article title and sections
\definecolor{color2}{RGB}{0,20,20} % Color of the boxes behind the abstract and headings

\usepackage{hyperref} % Required for hyperlinks
\hypersetup{hidelinks,colorlinks,breaklinks=true,urlcolor=color2,citecolor=color1,linkcolor=color1,bookmarksopen=false,pdftitle={Title},pdfauthor={Author},urlcolor=blue}

\usepackage{graphicx}
\usepackage[square]{natbib}
\usepackage{amsmath}
\usepackage{multirow}
\usepackage{subfigure}
\usepackage{pdflscape}
\usepackage[final]{pdfpages}

\usepackage{graphicx}
\usepackage{natbib}
\usepackage{amsmath}
\usepackage{amsfonts}
\usepackage{multirow}
\usepackage{subfigure}
\usepackage{tikz}
\usetikzlibrary{calc}

\usepackage{mathtools,amssymb,lipsum}

\usepackage{cuted}
\usepackage{boondox-calo}
\JournalInfo{Published in Physical Review Applied, 20, 044036 (2023)} % Journal information
\Archive{\href{https://doi.org/10.1103/PhysRevApplied.20.044036}{DOI: 10.1103/PhysRevApplied.20.044036}} % Additional notes (e.g. copyright, DOI, review/research article)

\PaperTitle{Magnetic levitation by rotation} % Article title

\Authors{Joachim Marco Hermansen$^{1,\dag}$, Frederik Laust Durhuus$^{2,\dag}$, Cathrine Frandsen$^2$, Marco Beleggia$^{3,4}$, Christian R. H. Bahl$^1$ and Rasmus Bjørk$^{1,*}$} % Authors
\affiliation{$^1$\textit{Department of Energy Conversion and Storage, Technical University of Denmark - DTU, DK-2800 Kgs. Lyngby, Denmark}} % Author affiliation
\affiliation{$^2$\textit{Department of Physics, Technical University of Denmark - DTU, DK-2800 Kgs. Lyngby, Denmark}} % Author affiliation
\affiliation{$^3$\textit{DTU Nanolab, Technical University of Denmark - DTU, DK-2800 Kgs. Lyngby, Denmark}} % Author affiliation
\affiliation{$^4$\textit{Department of Physics, Informatics and Mathematics, University of Modena and Reggio Emilia, 40125 Modena, Italy}} % Author affiliation
\affiliation{$\dag$Authors contributed equally} % Corresponding author
\affiliation{*\textbf{Corresponding author}: rabj@dtu.dk} % Corresponding author

\Keywords{} % Keywords - if you don't want any simply remove all the text between the curly brackets
\newcommand{\keywordname}{Keywords} % Defines the keywords heading name

\Abstract{A permanent magnet can be levitated simply by placing it in the vicinity of another permanent magnet that rotates in the order of 200 Hz.
This surprising effect can be easily reproduced in the lab with off-the-shelf components.
Here we investigate this novel type of magnetic levitation experimentally and clarify the underlying physics. Using a 19 mm diameter spherical NdFeB magnet as rotor magnet, we capture the detailed motion of levitating, spherical NdFeB magnets, denoted floater magnets, as well as the influence of rotation speed and magnet size on the levitation. We find that as levitation occurs, the floater magnet frequency-locks with the rotor magnet, and, noticeably, that the magnetization of the floater is oriented close to the axis of rotation and towards the like pole of the rotor magnet. This is in contrast to what might be expected by the laws of magnetostatics as the floater is observed to align its magnetization essentially perpendicular to the magnetic field of the rotor. Moreover, we find that the size of the floater has a clear influence on the levitation: the smaller the floater, the higher the rotor speed necessary to achieve levitation, and the further away the levitation point shifts. Despite the unexpected magnetic configuration during levitation, we verify that magnetostatic interactions between the rotating magnets are responsible for creating the equilibrium position of the floater. Hence, this type of magnetic levitation does not rely on gravity as a balancing force to achieve an equilibrium position. Based on theoretical arguments and a numerical model, we show that a constant, vertical field and eddy-current enhanced damping is sufficient to produce levitation from rest. This enables a gyroscopically stabilised counter-intuitive steady-state moment orientation, and the resulting magnetostatically stable, mid-air equilibrium point. The numerical model display the same trends with respect to rotation speed and the floater magnet size as seen in the experiments.}

\begin{document}

\flushbottom % Makes all text pages the same height

\maketitle % Print the title and abstract box

%\tableofcontents % Print the contents section

\thispagestyle{empty} % Removes page numbering from the first page

\section{Introduction}
Magnetic levitation is equally science fiction and present-day technology. Since Earnshaw's theorem prevents stable levitation with systems comprising only ferromagnets, current technologies such as Maglev trains \cite{hyung-woo_lee_review_2006}, flywheels \cite{supreeth_review_2022} and high speed machinery \cite{de_boeij_contactless_2009} rely on different physical compensation techniques to achieve levitation.

\begin{figure}
  \includegraphics[width=.50\textwidth]{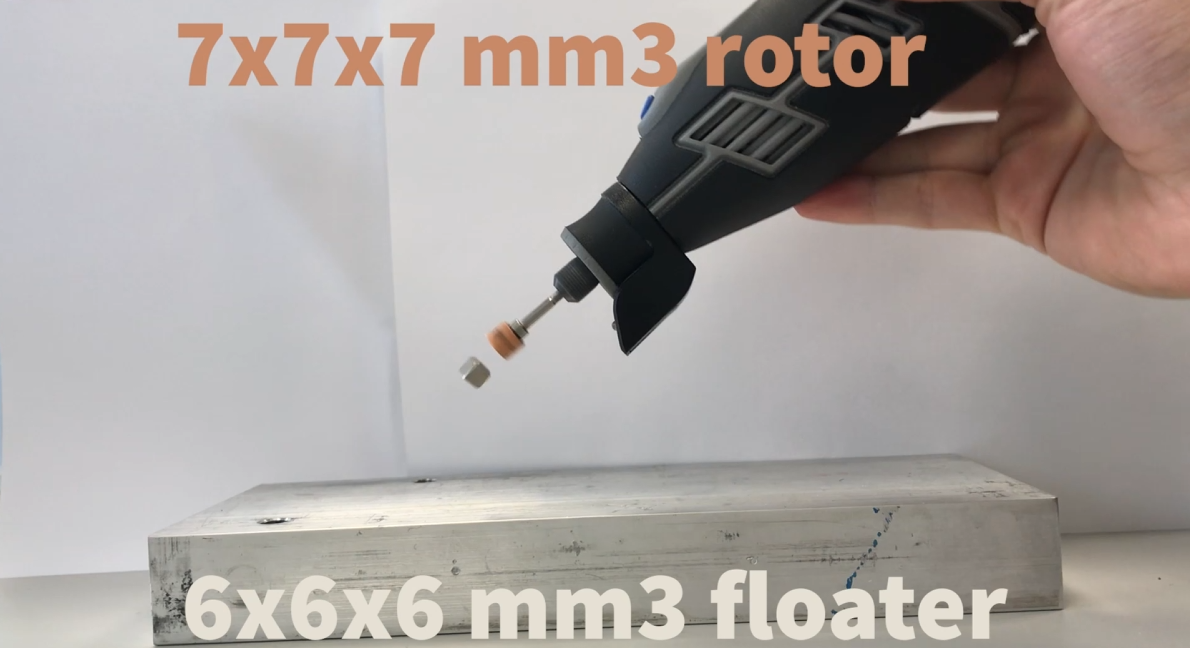}
  \caption{A demonstration of an easily reproducable experiment using a Dremel multitool to achieve magnetic levitation. Direct link: \href{https://youtu.be/o1eVSmnGJNc}{Video 1.}} \label{video_dremel}
\end{figure}

Recently, in 2021, a novel type of magnetic levitation was discovered \cite{ucar_polarity_2021} that uses two permanent magnets of similar size. One magnet, termed ``rotor'', mounted on a motor with its north and south poles oriented perpendicular to the rotation axis, is brought to rotate at angular velocities in the order of $10,000\;\text{RPM}$. The second magnet, termed ``floater'', is placed in the vicinity of the rotating magnet, it is spun in motion and levitates towards the rotor until it floats in space a few centimeters below it. The floater precesses with the same frequency as the rotor, and, if perturbed, experiences restoring forces that bring it back to its equilibrium position. It is quite surprising that magnetic levitation develops in such a relatively simple system: the magnetic forces do not suddenly create a stable minimum-energy point in space just by spinning one of the magnets, and yet levitation can be very easily reproduced in the lab with off-the-shelf components, as shown in Video \ref{video_dremel} or at Ref. \cite{YoutubeVideos}.

\renewcommand{\figurename}{Video}

Ferromagnetic levitation can generally be subdivided into three categories to counteract gravity. The first is active magnetic stabilisation, where a control system supplies the levitating magnet with an electromagnetic force to keep it balanced \cite{de_boeij_contactless_2009,jansen_magnetically_2008,gupta_applications_2011}. The second type is electrodynamic suspension, known from Maglev trains, where a moving magnet will induce a current in a stationary conductor producing a repulsive force that increases with the speed of the moving magnet \cite{hyung-woo_lee_review_2006}. Finally, the last category is spin stabilised levitation, where the levitating object is spinning and uses the gyroscopic effect to keep the system stable. This is the effect seen in the Levitron \cite{michaelis_stability_2015,michaelis_horizontal_2014}, and the magnetic Paul trap\cite{perdriat_planar_2022,sackett_magnetic_1993}.

The Levitrons spinning rate is typically $500\;\text{RPM}$ \cite{gans_dynamics_1998}, much lower than seen in the novel type of magnetic levitation. Furthermore, since it is not a driven system, once friction with the air slows the magnet spinning rate, it loses levitation. The magnetic Paul trap uses a rotating gradient field for levitation, hence is driven, however it relies on a balance between gravity and magnetic repulsion for vertical stability. For the levitation studied in this paper, both attraction and repulsion are magnetic, so the floater magnet is fixed relative to the rotor even when moving and rotating the whole device. Thus the phenomenon can be used for the trapping and 3-dimensional, contactless manipulation of magnetic objects, similar to how Refs.\ \cite{pham_dexterous_2021,tabor_adaptive_2022,dalton_attracting_2022} used spinning dipole magnets for handling metal spheres. Ref.\ \cite{pham_dexterous_2021} relies on eddy currents for inducing a coupling and complex feedback control for stability. For the levitation presented here, eddy currents are relevant but not required for levitation and the phenomenon is inherently stable.
Determining the full range of applications requires a deeper understanding of the present phenomenon, in particular the scalability and range of stability for levitation, but other potential applications could include trapping and manipulating of ferromagnetic microparticles.

In the recent pioneering work \cite{ucar_polarity_2021}, H. Ucar explains the novel magnetic levitation in terms of two key concepts: polarity free magnetic repulsion (PFR) and magnetic bound state (MBS). PFR is a short-range repulsive force with $F\propto \frac{1}{r^{7}}$ decay with distance that is postulated to originate from the synchronisation of the floater magnet and the rotor magnet. It is assumed that the rotor and floater will have a constant phase relative to each other and from that assumption the repelling force emerges. However, this assumption is not clearly validated in the work. MBS is the stable levitation that can occur if the rotor magnet is tilted and/or shifted with respect to the rotation axis as this in combination with the tilt of the floater magnet can result in an attractive force with a stable equilibrium point. However, this is not rigorously and systematically investigated in H. Ucar's work.

The pioneering paper by H. Ucar, and a short follow-up manuscript comparing the system to a Kapitza pendulum \cite{Ucar_2023}, does not decipher many key aspects of the levitation mechanism. It is not discussed how stable the magnetic levitation phenomenon is and what kinds of instabilities can cause levitation to cease nor when these occur. Furthermore the influence of rotor speed, floater size and floater magnetisation on levitation is not discussed. The detailed orientation of the floater relative to the rotor is also not quantified experimentally. Furthermore, on the theoretical side, H. Ucar exclusively considered energy conserving models of magnetostatic coupling, and an analysis of eddy currents and the effect of drag is not considered. Studying systematically how the various parameters influence the levitation is crucial to understanding the physics behind this novel levitation effect. In this paper, we consider these aspects experimentally using high-speed video tracking. We then discuss the various physical effects in the system and their relevance. Finally, we simulate a simple dynamical model of the system to elucidate the mechanisms behind levitation.

\section{Experimental setup}
%To systematically investigate these aspects, w
We realized an experimental setup as follows. A spherical NdFeB rotor magnet with a diameter of $19 \;\text{mm}$ and a nominal remanence of 1.22-1.26 T was fixed using epoxy into a 3D printed plastic holder. The rotor orientation was specified and determined by placing a neodymium magnet on the side of the 3D printed rotor mount, such that the magnetization direction was kept aligned while the epoxy cured. The holder was subsequently mounted onto the shaft of a high speed motor (Vevor JST-JGF-F65A) with speed control, assembled on an aluminium support. The motor allows for experiments at rotational speeds up to $400 \;\text{Hz}$ or $24,000 \;\text{RPM}$, calibrated using an induction coil and oscilloscope setup. This is comparable to the work of H. Ucar, who uses rotor speeds between $5,530\;\text{RPM}$ and $105,000\;\text{RPM}$ \cite{ucar_polarity_2021}.  The intrinsic alternating magnetic field of the motor at a typical distance of the floater magnet is assumed to be negligible. The experimental setup is shown in Fig. \ref{Fig.ExpSetup}.

After epoxying the rotor magnet onto its holder, all three spatial components of the magnetic field produced by the rotor magnet were measured by a Hall probe in 361 positions in a 190 mm by 190 mm slice 155 mm above the center of the rotor magnet. Fitting the measured field with a dipolar field resulted in a polar angle $\theta_r=90.5^\circ\pm2.5^\circ$ (cf. Fig. \ref{Fig.Angles}) i.e. close to the desired direction.

\renewcommand{\thefigure}{1}
\renewcommand{\figurename}{Fig.}
\begin{figure}[t]
\begin{center}
\includegraphics[width=.50\textwidth]{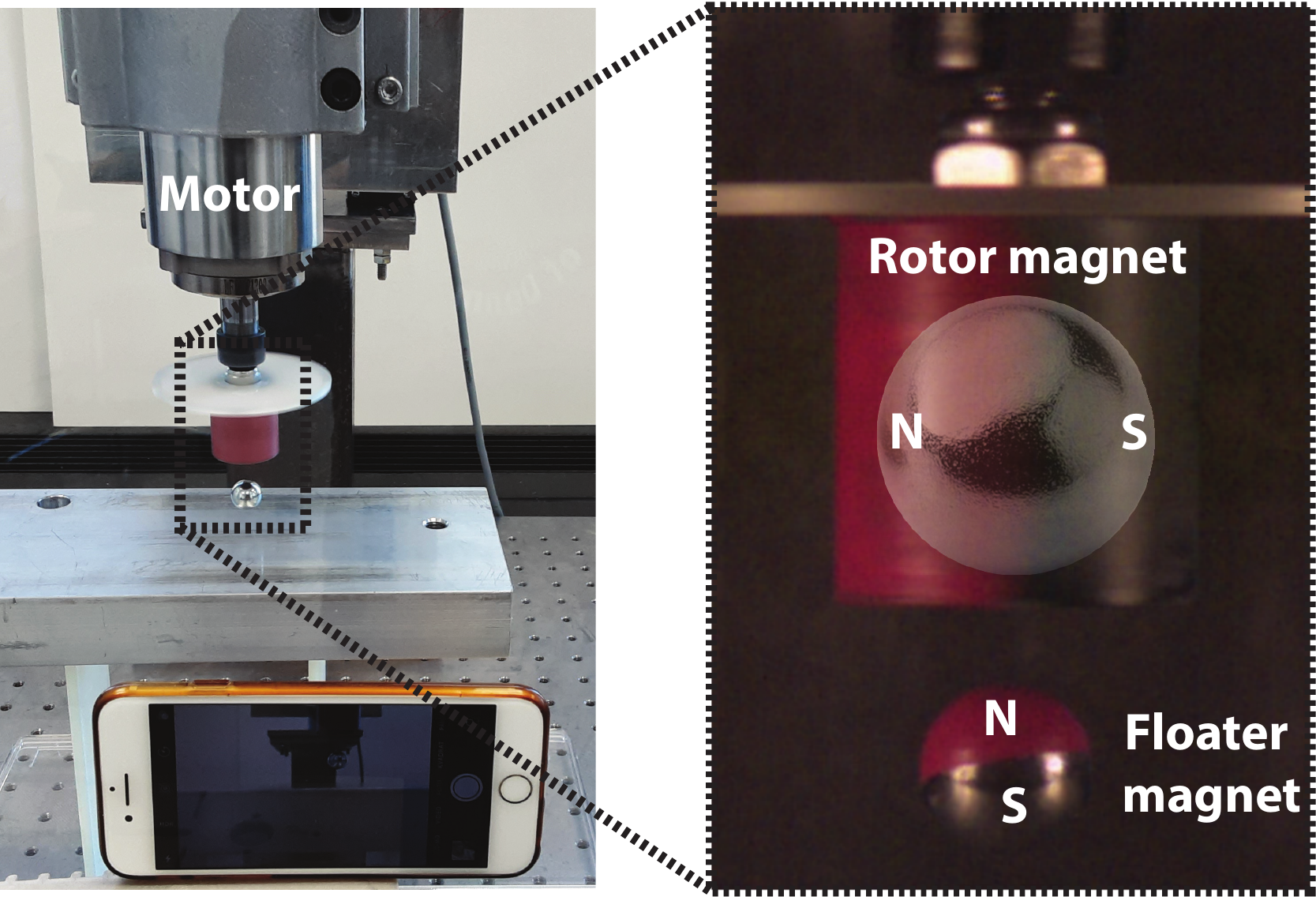}
\caption{The experimental setup, including a closeup of the rotor and floater magnet. The closeup is an image taken with the high-speed camera and where the floater magnet has been painted to indicate its magnetic poles. The floater magnet can clearly be seen to be levitating.}
\label{Fig.ExpSetup}
\end{center}
\end{figure}

During experiments the dynamical behaviour of the floater magnet was filmed and the recordings post-processed with the motion-tracking software Tracker to determine the position and orientation of the floater as a function of time, as demonstrated in Video \ref{VideoTracker}. Two different camera setups were used. The low-speed option consisted of an iPhone 7 camera recording at 30 frames per second. For faster dynamics a Chronos 1.4 high speed camera recording at 1057 frames per second was used.

\renewcommand{\thefigure}{2}
\renewcommand{\figurename}{Video}
\begin{figure}
  \includegraphics[width=.50\textwidth]{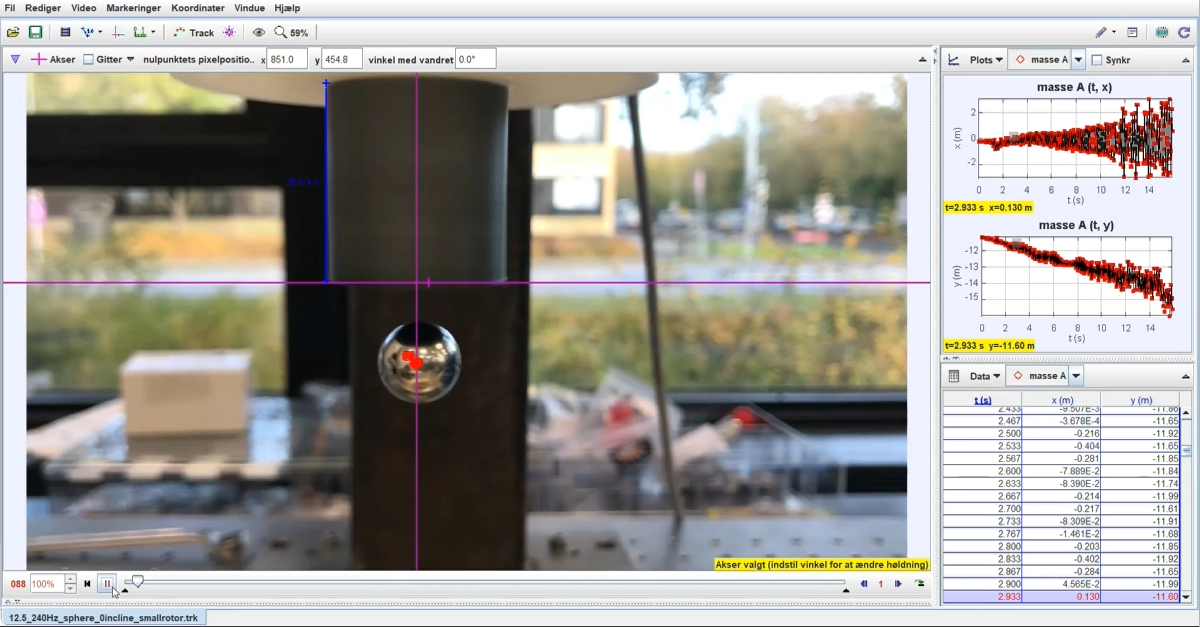}
  \caption{A demonstration of the motion-tracking software Tracker used to determine the position and orientation of the floater as a function of time. Direct link: \href{https://youtu.be/agfOx3xIOWE}{Video 2.}}\label{VideoTracker}
\end{figure}

To consistently reproduce and test the magnetic levitation phenomenon, we adopted the following experimental protocol. First, the motor was spun at the desired speed with a 30 mm thick aluminium plate placed 35 mm below the rotor magnet. The purpose of the aluminium plate is to dampen the initial transient motion of the floater through eddy currents, allowing for an easier initial levitation. After being inserted manually  beneath the rotor, the floater magnet quickly levitates towards its equilibrium position.  As soon as this is reached, the aluminium plate is removed and it is thus only present during the initial levitation of the floater magnet. We note that with practice the floater magnet can be levitated without the aluminium plate. The procedure described above is shown on a video recording in Video \ref{Video1} or at Ref. \cite{YoutubeVideos}.

\renewcommand{\thefigure}{3}
\renewcommand{\figurename}{Video}
\begin{figure}
  \includegraphics[width=.50\textwidth]{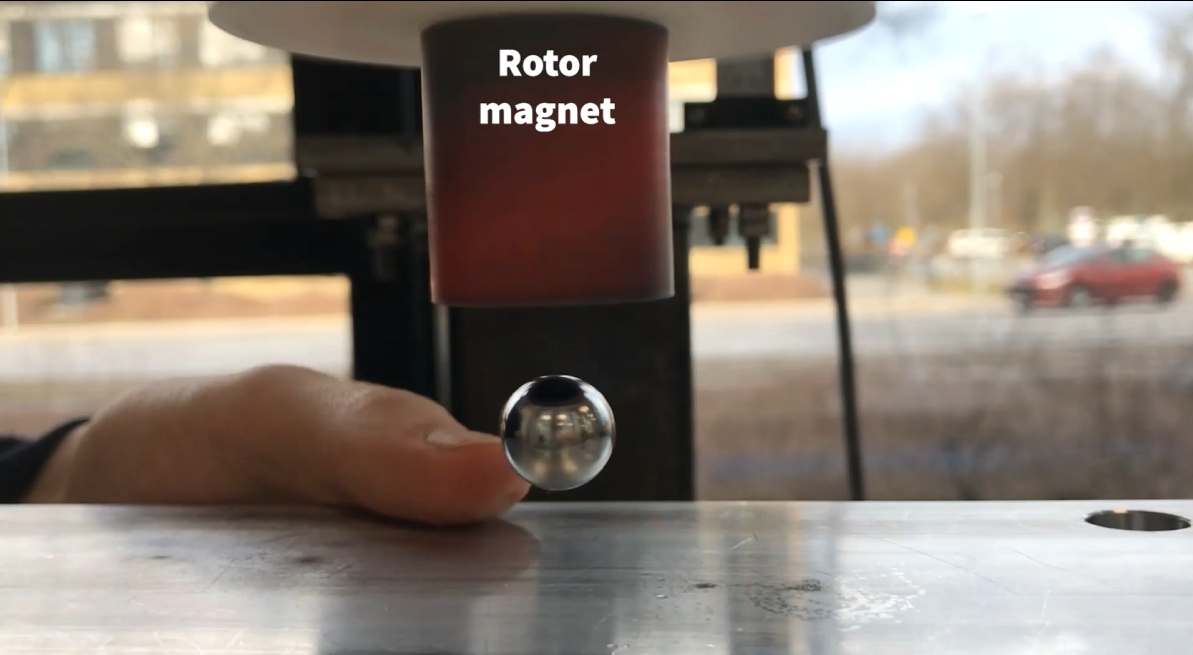}
  \caption{The procedure to reproduce the magnetic levitation phenomenon. Direct link: \href{https://youtu.be/5zHKNOcsgeQ}{Video 3.}}\label{Video1}
\end{figure}

\section{Results}
The magnetic levitation was investigated by three different experiments. The first was aimed at understanding the detailed alignment of the rotor and floater magnets for a fixed set of experimental parameters. The second focused on the dynamics of the floater magnet as a function of rotor speed for a fixed-size floater. The third experiment investigated the role of rotor size and magnetization on levitation.

All data mentioned below, including all video captures, are available from the data repository, Ref. \cite{Data_2023}.

\subsection{Floater alignment}

\renewcommand{\thefigure}{2}
\renewcommand{\figurename}{Fig.}
\begin{figure}[!t]
\begin{center}
\includegraphics[width=.30\textwidth]{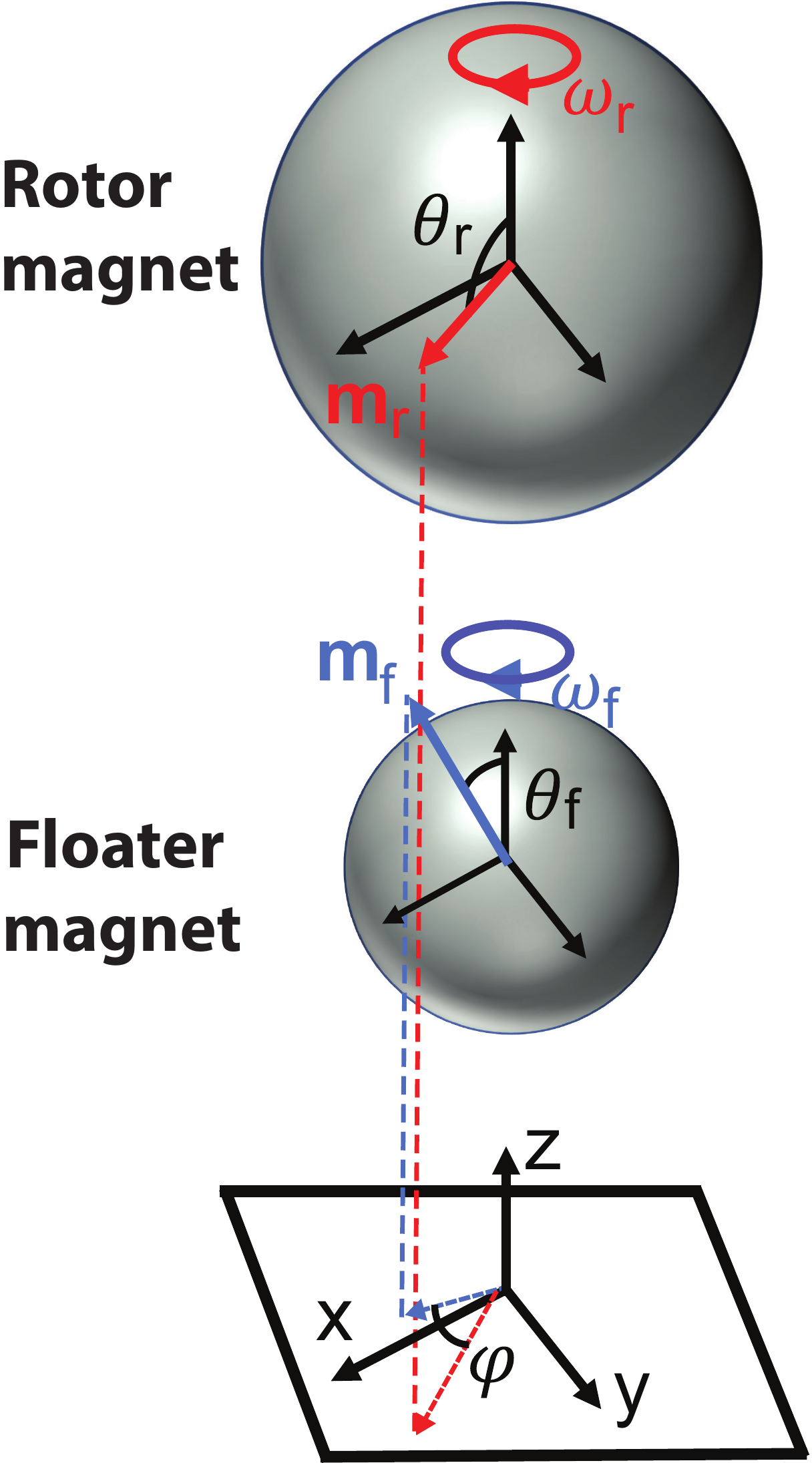}
\caption{The phase angle, $\varphi$, between the floater and the rotor magnets. The phase angle is the angle between the projection onto the $xy$-plane of the respective magnetization vectors of the two magnets, m$_r$ and m$_f$. The levitation distance is the center-to-center distance between the rotor and the floater magnets.}
\label{Fig.Angles}
\end{center}
\end{figure}

First we establish the exact movement of the floater magnet. It is particularly important to determine the orientation of the floater magnet, i.e. the direction of its magnetization during rotation. To establish the polar angle of the floater magnet, $\theta_f$, as well as the phase angle, $\varphi$, between the rotor and floater magnetization vectors in the $xy$-plane (cf. Fig. \ref{Fig.Angles}), an experiment was performed with a spherical NdFeB floater magnet with a diameter of 12.7 mm and a remanence of 1290-1320 mT, with the rotor magnet rotating at 200 Hz. The rotor magnet was painted red on its magnetic northern hemisphere and the floater magnet was painted red on its northern pole side, to allow their orientations to be tracked. The details showing the tracking method can be found in the supplementary material \cite{Supplementary_material} and an example experimental video is shown in Video \ref{Video_High_speed_camera_tracking_at_1057_fps}. Noticeably, the polar angle is determined to be $\theta_f=7^{\circ}\pm4^{\circ}$ and the phase angle $\varphi=6.4^{\circ}\pm5.1^{\circ}$. This configuration (seen in Fig. \ref{Fig.ExpSetup}) is surprising from a purely magnetostatic point of view as the magnetization of the floater magnet is almost vertical and makes a North-North orientation with the rotor magnet, i.e. it points essentially perpendicular to the magnetic field of the rotor magnet. We stress that the orientation of the magnetization vectors shown in Fig. \ref{Fig.Angles} may equally likely be North-North, as South-South.

\renewcommand{\thefigure}{4}
\renewcommand{\figurename}{Video}
\begin{figure}
  \includegraphics[width=.40\textwidth]{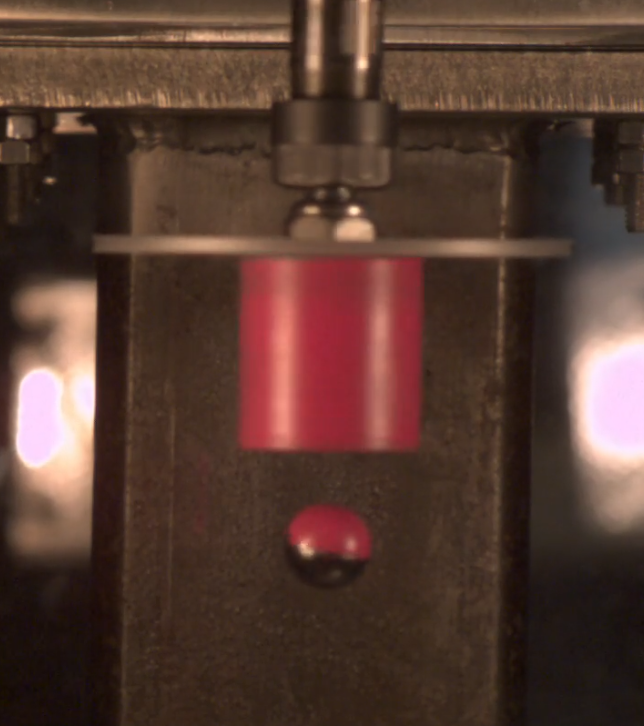}
  \caption{High speed camera footage at 1057 frame per second of the painted magnet levitating. The rotor magnet is rotating at 200 Hz. Direct link: \href{https://youtu.be/VXfhQ_TIChY}{Video 4.}}\label{Video_High_speed_camera_tracking_at_1057_fps}
\end{figure}

\subsection{Floater dynamics}
To establish the floater movement dynamics a series of experiments using the 12.7 mm diameter spherical NdFeB floater magnet with a remanence of 1290-1320 mT was conducted. The rotor speed was varied in steps of 5 Hz from 130 Hz to 280 Hz, with an additional measurement at 142.5 Hz, as it is the lowest speed for which levitation was possible. Depending on the dynamics 2-5 experiments were done for each rotor speed.

For each experiment the dynamics of the floater magnet was tracked using the low-speed video setup. The recorded videos reveal one semi-stable behavior and four different types of vibrational modes of the floater magnet, as a function of the frequency of the rotor magnet. The semi-stable behavior is characterized by rotation during levitation without additional oscillation modes i.e. displacement amplitudes of $<$1 mm on timescales shorter than 0.5 s. The four vibrational modes are an Up-down mode with oscillations up to 4 mm, defined by vertical oscillations of the floater magnet, a Side mode with oscillations up to 8 mm, where the floater magnet oscillations are in the horizontal plane and swing out in a horizontal circular path around the origin, a Mixed mode with oscillations up to 3.5 mm and 7 mm for the Up-down mode and Side modes respectively and a U-shaped mode with oscillations up to 7.5 mm, where the floater oscillations have a clear U-shape. Each of these modes is illustrated in videos \ref{Video_semi}, \ref{Video_updown}, \ref{Video_side}, \ref{Video_mixed} and \ref{Video_U} or at Ref. \cite{YoutubeVideos}.

In all experiments the levitation is unstable and eventually the floater magnet drops away from the rotor magnet, if the motion of the floater is not damped by eddy current generated in e.g. an aluminium plate below the floater.

%\begin{figure}[t]
%\begin{center}
%\includegraphics[width=.50\textwidth]{Modes_with_mixed}
%\caption{An illustration of the different modal movements of the floater magnet, as labelled in the individual figures. An animation showing the movements are available in the video supplementary material.}
%\label{Fig.ModesIllustrated}
%\end{center}
%\end{figure}
%\renewcommand{\thefigure}{5}
\setcounter{figure}{4}
\renewcommand{\thefigure}{\arabic{figure}}
\renewcommand{\figurename}{Video}
\begin{figure*}[!htb]
    \centering
    \begin{minipage}{.19\textwidth}
        \centering
        \includegraphics[width=0.98\textwidth]{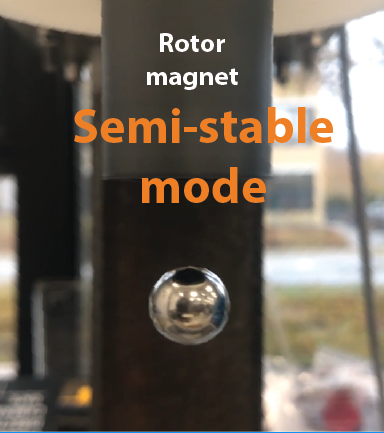}
        \caption{The semi-stable  mode. Direct link: \href{https://youtu.be/7w2LRoC8qYk}{Video 5.}}
        \label{Video_semi}
    \end{minipage}%
    \begin{minipage}{0.19\textwidth}
        \centering
        \includegraphics[width=0.98\textwidth]{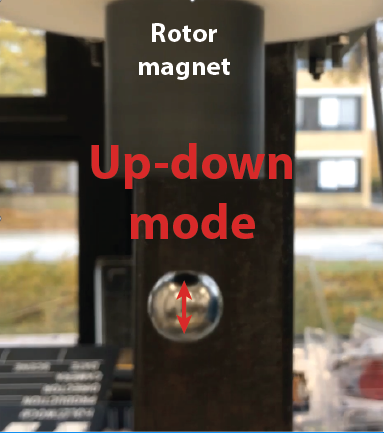}
        \caption{The up-down movement mode. Direct link: \href{https://youtu.be/wfkzfzMpaqI}{Video 6.}}
        \label{Video_updown}
    \end{minipage}
    \begin{minipage}{0.19\textwidth}
        \centering
        \includegraphics[width=0.98\textwidth]{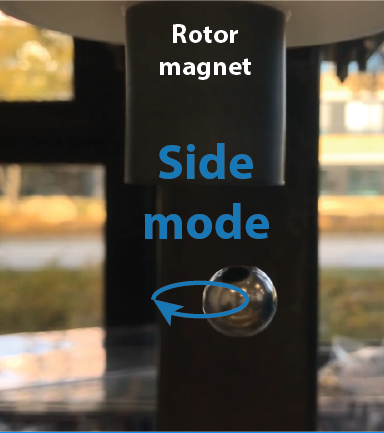}
        \caption{The side movement mode. Direct link: \href{https://youtu.be/FnofhF7RTg4}{Video 7.}}
        \label{Video_side}
    \end{minipage}
    \begin{minipage}{0.19\textwidth}
        \centering
        \includegraphics[width=0.98\textwidth]{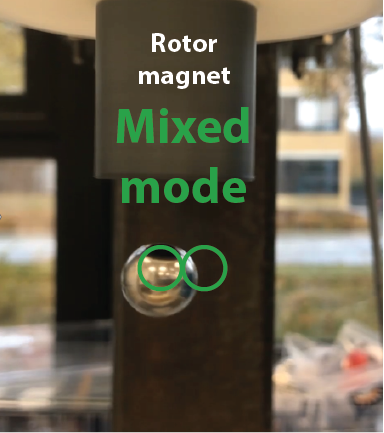}
        \caption{The mixed movement mode. Direct link: \href{https://youtu.be/TyiaT4J5iD0}{Video 8.}}
        \label{Video_mixed}
    \end{minipage}
    \begin{minipage}{0.19\textwidth}
        \centering
        \includegraphics[width=0.98\textwidth]{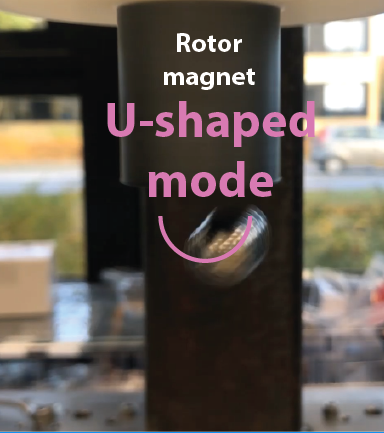}
        \caption{The U-shaped movement mode. Direct link: \href{https://youtu.be/PMXhLOYt6tA}{Video 9.}}
        \label{Video_U}
    \end{minipage}
\end{figure*}

From the tracked data the levitation time and initial levitation distance is measured (Fig. \ref{Fig.ModesQuantified}a  and \ref{Fig.ModesQuantified}b respectively). In Fig. \ref{Fig.ModesQuantified}a, it can be seen that the dynamics of the system is highly dependent on the rotor speed. Below 142.5 Hz no levitation was observed. At low speeds between 142.5 Hz and 180 Hz there are no vibrational modes present and the floater magnet slowly descends until finally dropping. At a rotor speed of 185 Hz the dynamics suddenly change and an Up-down motion is present, which grows with time resulting in an increased instability and a resulting decrease in levitation time. At 195 Hz the dynamics change once again and the Side mode appears. With this mode a sudden increase in levitation time was observed. Increasing the rotor speed further results in a rapid drop in levitation time. At 215 Hz a new mode appears, with complex dynamics. The floater motion is here a mix of the Side and Up-down modes.  Finally, at even higher speeds the U-shaped mode appears. The highest speed at which the levitation could occur was 280 Hz, beyond that speed the rotor and floater magnets collide, and the floater magnet is flung off.

\renewcommand{\thefigure}{3}
\renewcommand{\figurename}{Fig.}
\begin{figure}[!t]
    \subfigure{\includegraphics[width=0.49\textwidth]{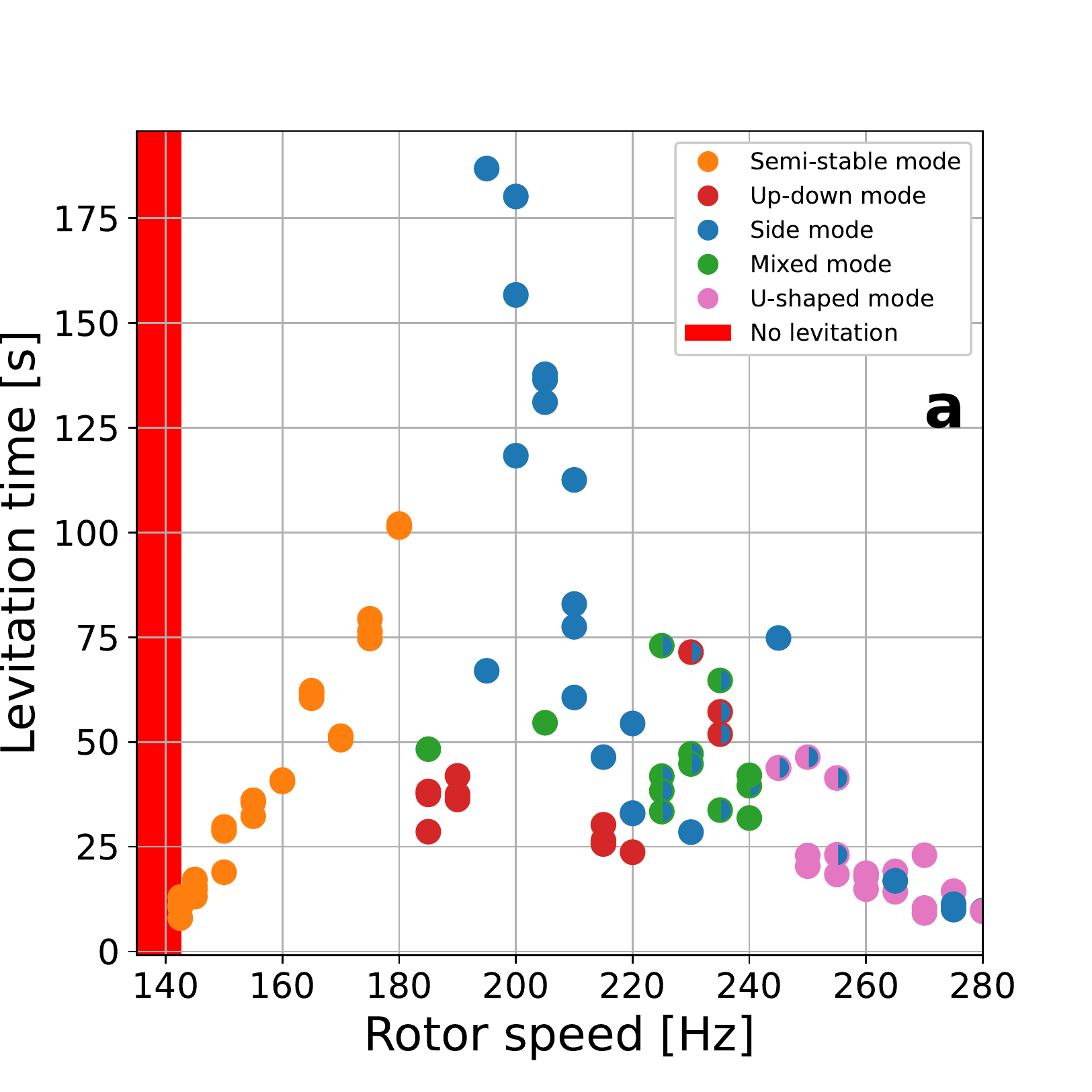}}
    \subfigure{\includegraphics[width=0.49\textwidth]{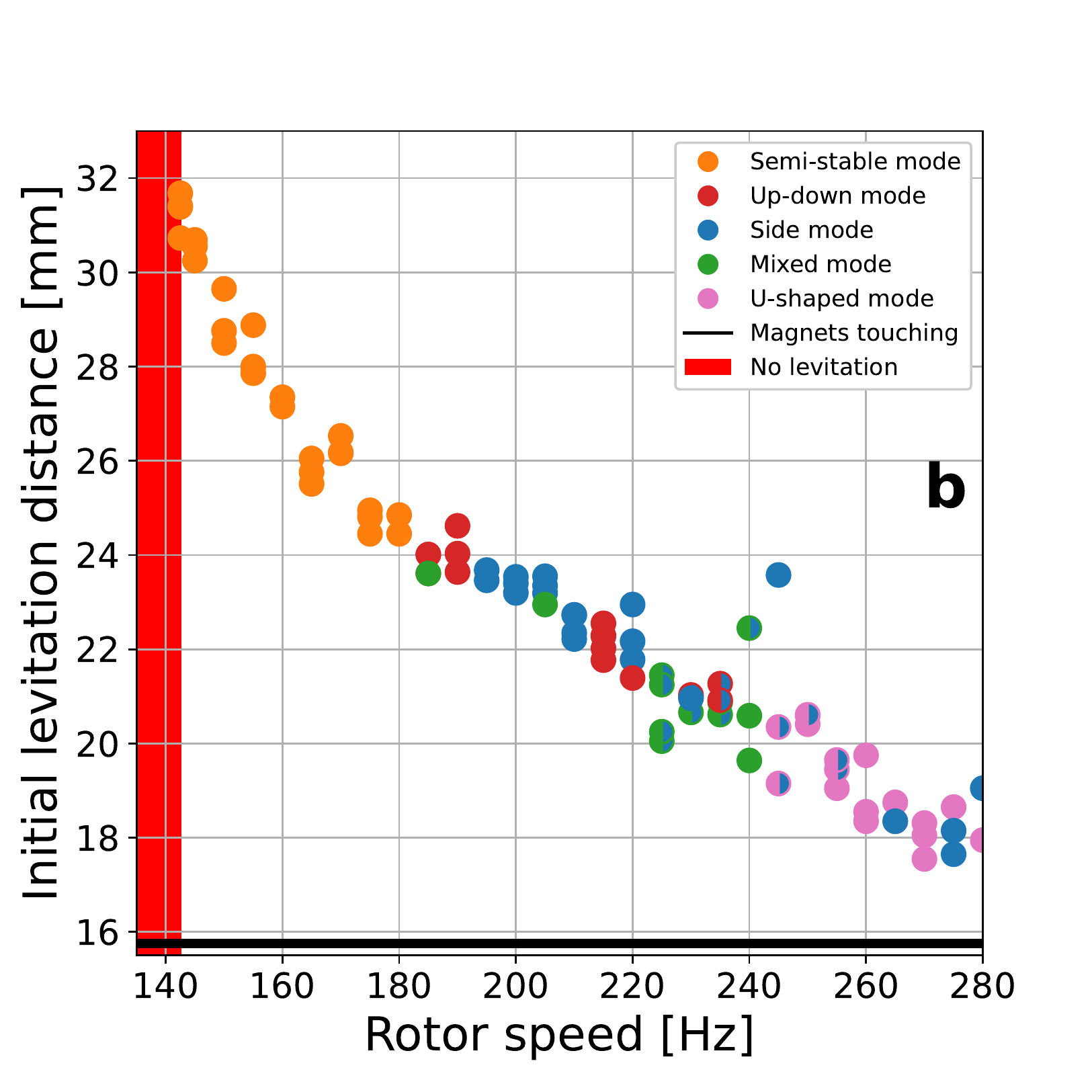}}
    \caption{\textbf{a}) Levitation time as function of rotor speed and \textbf{b}) initial levitation distance as function or rotor speed. For both figures the color of the dots denote the mode occurring during levitation. The multicolored dots mean that the mode changes from the left colored mode to the right colored mode, e.g. for the red/blue case the mode changes from an Up-down mode exclusively to a Side mode exclusively. The levitation distance is defined as the distance from the center of the rotor spherical magnet to the center of the floater spherical magnet.}
        \label{Fig.ModesQuantified}
\end{figure}

Regarding the initial levitation distance, shown in Fig. \ref{Fig.ModesQuantified}b, it is clear that regardless of vibrational mode, the initial levitation distance becomes shorter at higher rotor speeds. This clearly shows that the force between the two magnets changes with rotation speed. Note that it is the initial levitation distance that is shown. For all rotor speeds for this size floater magnet, the levitation is unstable, unless e.g. an aluminium plate is used to dampen the motion, and thus the levitation distance increases over time. If we specifically consider the semi-stable regime where no oscillating modes are present, the fall rate can be investigated in more detail. Here, for each experiment between 142.5 Hz and 180 Hz, a linear curve was fitted to the dynamics of the floater, to find the fall rate. The time dynamics are available in the data repository and the determined falling rates are shown in Fig. \ref{Fig.FallingRate}. As can be seen the falling rate decreases with increased rotor speed.

\renewcommand{\thefigure}{4}
\renewcommand{\figurename}{Fig.}
\begin{figure}[t]
\begin{center}
\includegraphics[width=.50\textwidth]{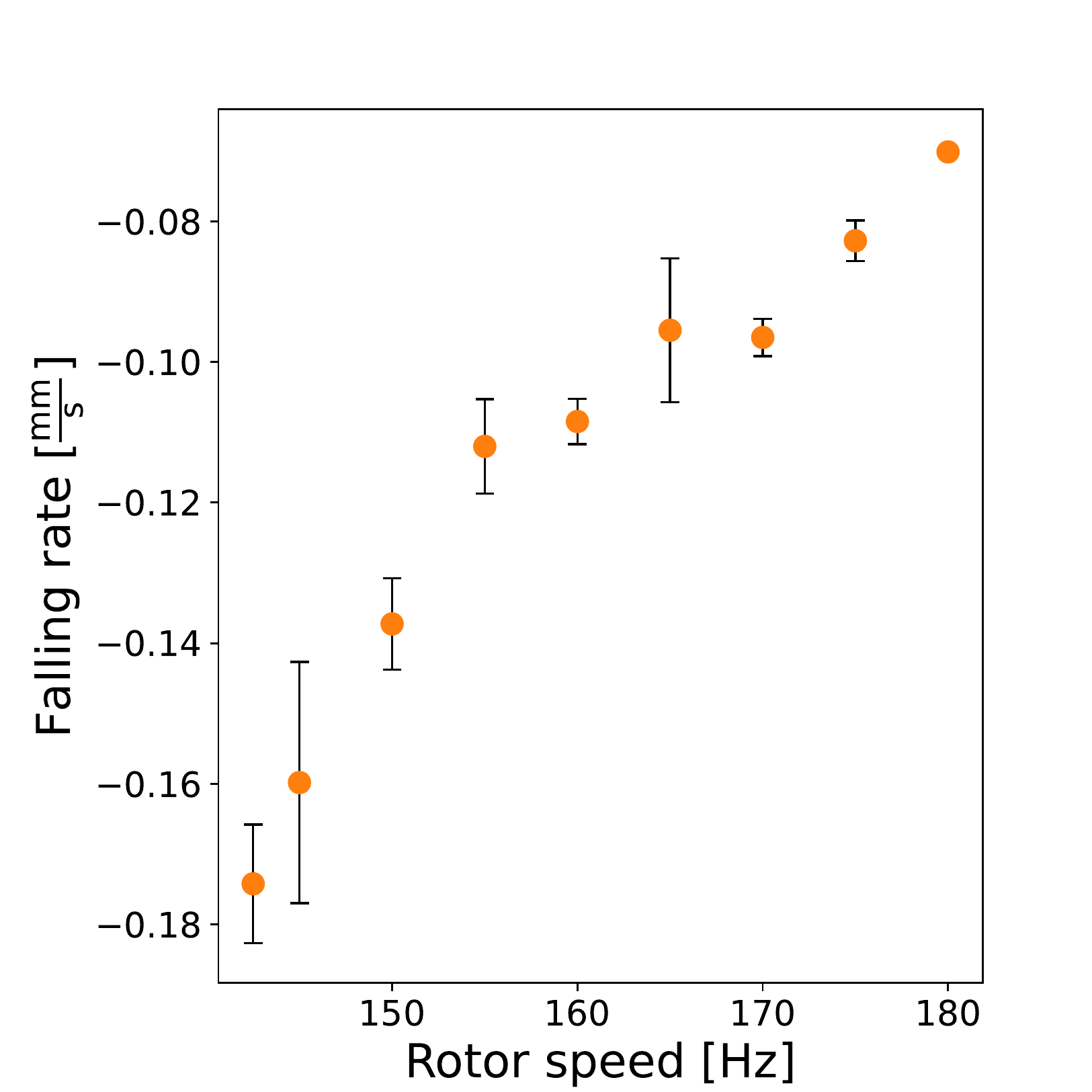}
\caption{The fall rate of the floater magnet as function of motor speed for the semi-stable region. The errorbars are the standard deviation of three experiments for each rotor speed.}
\label{Fig.FallingRate}
\end{center}
\end{figure}

\subsection{Floater magnetization and size}
It is also of interest to investigate the influence of floater size and magnetization on the rotation dynamics. The floater size is changed by using differently sized floater magnets. These magnets have a slightly varying magnetization, with the 5 mm and the 12.7 mm sphere having a remanence of 1290-1320 mT, the 6 mm, 8 mm, 19 mm and 26 mm spheres have a remanence of 1220-1260 mT and the 10 mm and 30 mm spheres have 1260-1290 mT. The weight of the spheres are as follows: the 5 mm sphere weighs 0.5 g, the 6 mm 0.86 g, the 8 mm 2 g, the 10 mm 4 g, the 12.7 mm 8.2 g, the 19 mm 27 g, the 26 mm 70 g, and the 30 mm 110 g.

To vary the remanence, heating was used to reduce the magnetization of six spherical magnets with diameter of 12.7 mm. The magnets were heated in an oven for 1 hour each at temperatures varying from $150^\circ $C to $200^\circ $C with $10^\circ $C intervals, in order to gradually demagnetise them \cite{haavisto_magnetic_2014}. Along with each spherical magnet, a cube magnet of size $7\times7\times7$ mm$^3$ and with the same remanence was placed in the oven. Following heat treatment, the magnetization of the cube magnet was measured on a Brockhaus Hystograph HG 200, which requires samples with flat surfaces, hence the use of the cube magnets. To avoid remagnetizing the magnet in the process the field applied was $\pm0.5\;\mathrm{kA}/\mathrm{m}$ during the measurement. The spherical magnets were then assumed to have sustained the same loss in magnetization.

We first consider the minimum rotor speed needed to achieve levitation. Here levitation is defined as the floater magnet levitating for more than 0.5 s. The minimum rotor speed as function of both the floater diameter and the remanence of the floater magnet is shown in Fig. \ref{Fig.LevDistSizeMag}a. To determine the minimum rotor speed, the speed of the rotor was decreased from 400 Hz in steps of 2.5 Hz for the floater magnet diameter variation and in steps of 1 Hz for the remanence variation. The data points shown in the figure is the lowest rotor speed where levitation was possible. As can be seen, the magnetization does not influence the minimum rotor speed, as decreasing the magnetisation by a third from $1180\;\text{mT}$ to $760\;\text{mT}$ only reduces the minimum rotor speed minimally. However, the sphere size has a very clear influence, with the smaller the floater magnet, the higher the rotor speed necessary to achieve levitation. This thus makes it clear that the volume of the floater is important in the dynamics, most likely because inertial effects are playing a dominating part in levitation although other size effects such as eddy currents could potentially also be present.

We also consider the initial levitation distance as function of both the floater magnet spheres diameter and the remanence of the floater magnet. This is shown in Fig. \ref{Fig.LevDistSizeMag}b. As can be seen from the figure, an increased remanence results in a slightly increased levitation distance. Oppositely, increasing the size of the floater magnet results in a shorter levitation distance. This again hints at the interplay between inertial effects and (electro)magnetic forces. Note that the experiments are not done at the same rotor speed, as the floater magnet diameter variation needs to be done at high rotor speeds to enable all tested spherical magnets to levitate.

\renewcommand{\thefigure}{5}
\renewcommand{\figurename}{Fig.}
\begin{figure*}[t]
    \subfigure{\includegraphics[width=0.49\textwidth]{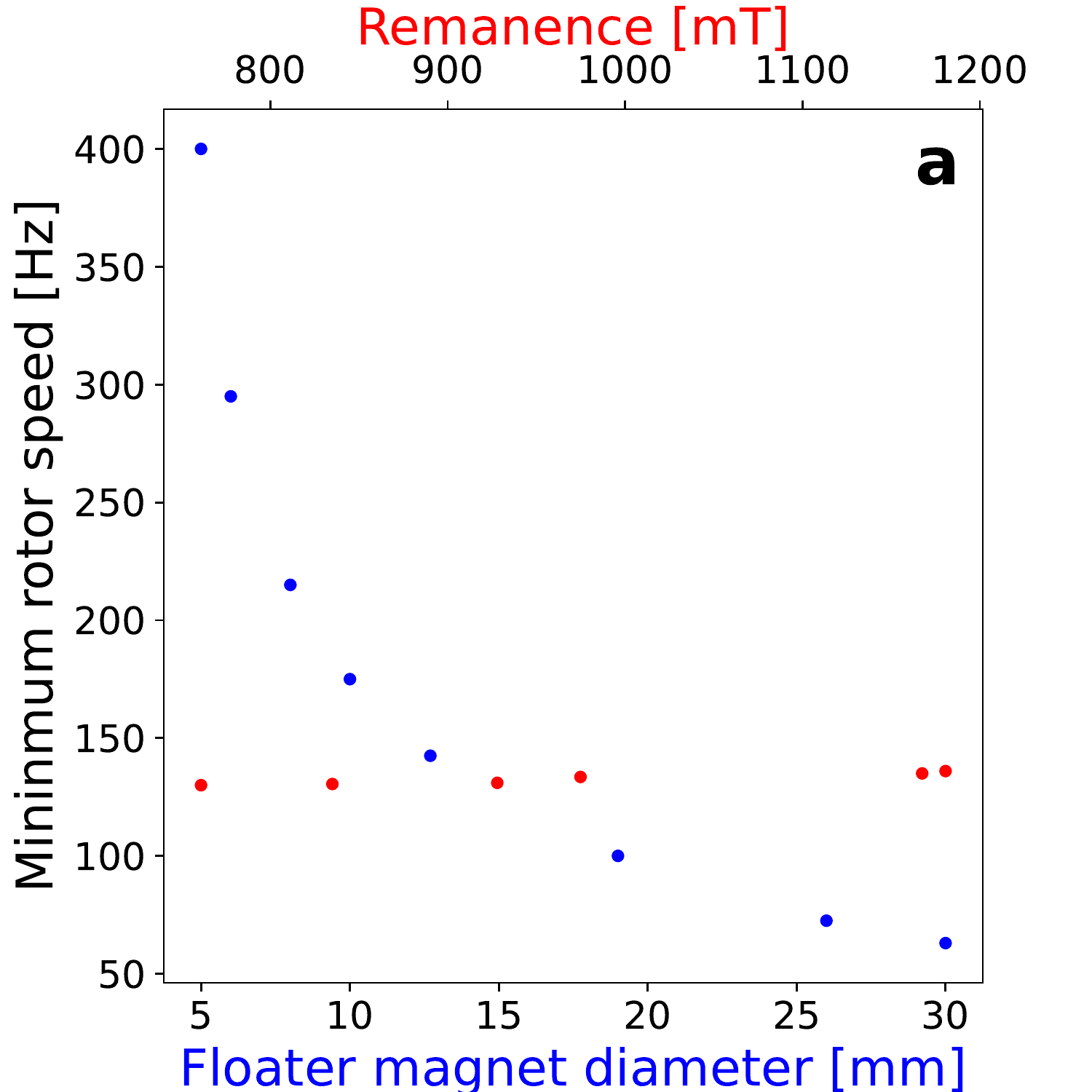}}
    \subfigure{\includegraphics[width=0.49\textwidth]{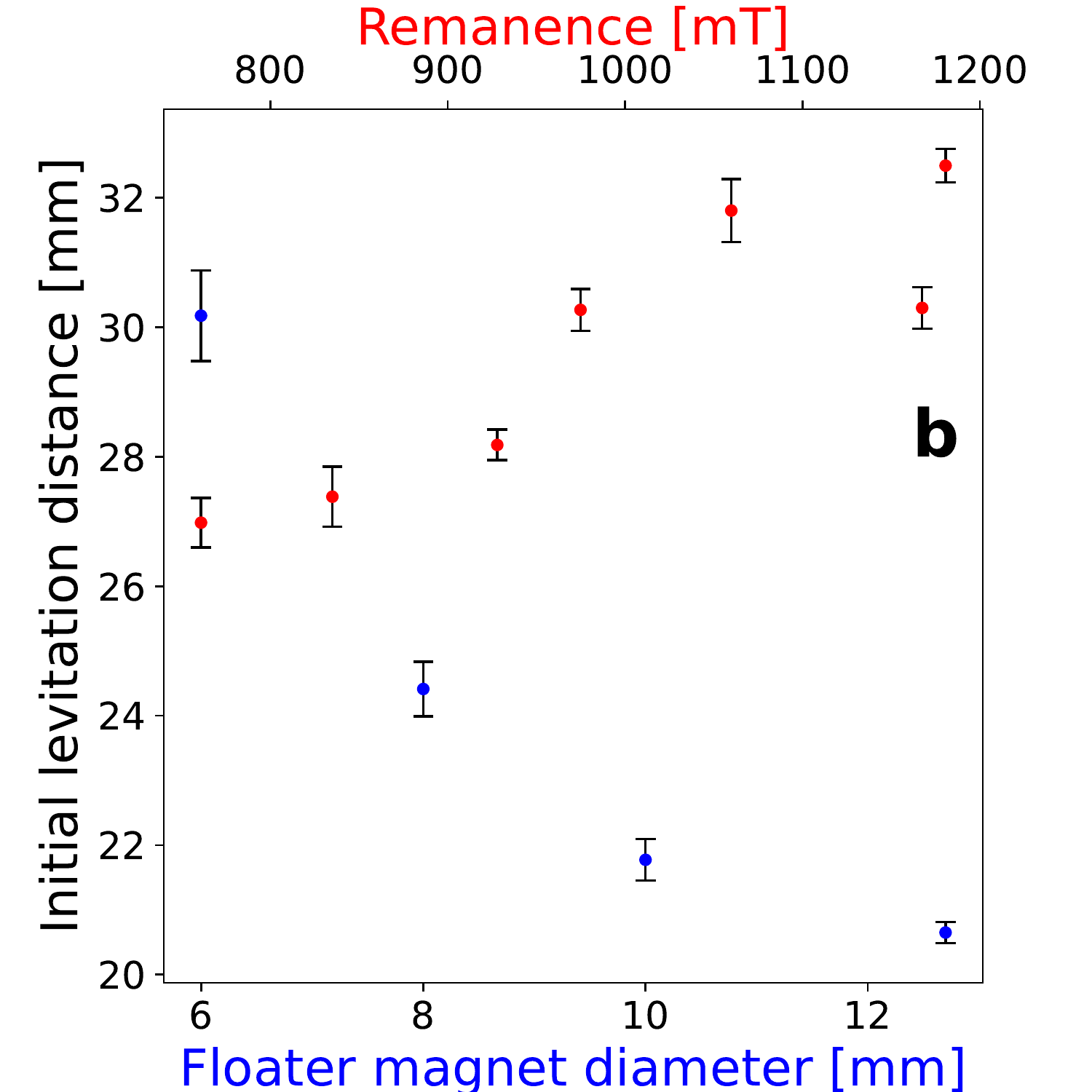}}
    \caption{\textbf{a}) Minimum rotation speed to achieve levitation as function of floater magnet spheres diameter (blue points, bottom x-axis) and remanence (red points, top x-axis). \textbf{b}) The initial levitation distance as function of floater magnet spheres diameter (blue points, bottom x-axis) and remanence (red points, top x-axis) at a rotor speed of 150 Hz for the remanence variation and 300 Hz for the floater magnet diameter variation. The error bars on b) given are the standard deviation of the position as determined from the tracker software.}
        \label{Fig.LevDistSizeMag}
\end{figure*}

%\begin{figure}[t]
%\begin{center}
%\includegraphics[width=.50\textwidth]{Tilt_levidist}
%\caption{Average tilt of the floater magnet at different levitation distances.}
%\label{Fig.Tilt}
%\end{center}
%\end{figure}

\subsection{Discussion}
In all measurements presented above, the levitation is limited in time. However, if an aluminium plate is left below the floater, to dampen its oscillation modes by eddy current dampening, levitation will occur perpetually. In the supplementary material the position of a 12.7 mm floater with an aluminium plate 33.5 mm below the centre of the rotor magnet and a rotation rate of 240 Hz is shown \cite{Supplementary_material}. Initially, the floater magnet falls with a rate of 0.3 mm/s for the first 20 seconds, after which is reaches an equilibrium height where it levitates for 40 minutes without any signs of stopping.

Additional experiments were done to explore the nature of the levitation phenomenon. First, it was verified that if a soft-magnetic steel sphere was used as floater, it could not be made to levitate. This was verified for steel spheres of diameters 8 mm, 10 mm, 12.7 mm and 20 mm and for rotor speeds in the range 50-400 Hz. Regardless of size and rotor speeds, the steel sphere is always attracted to the rotor magnet, i.e. there is no repulsion force in the system.

We also performed an experiment, where the floating magnet was an epoxy-bonded cylindrical-shaped magnet with 77.5 vol\% NdFeB magnet material. At 200 Hz this could also be made to levitate.

We wish to note that the magnetic levitation phenomenon does not require a complicated experimental setup to be realized. We have successfully demonstrated the phenomenon using a simple block magnet glued to a bolt mounted to a high-speed Dremel cutting tool, which can then be used to levitate another block magnet. A video of this is shown in Video \ref{video_dremel} or at Ref. \cite{YoutubeVideos}.

\section{The physics of levitation \label{sec:The_physics_of_levitaion}}

There are a number of different experimental observations to explain, including but not limited to: translational stability, frequency locking of rotor and floater rotation, and the relative direction of the magnetic moments. Below we summarise the relevant physical effects in the system and show how none of the known mechanisms of magnetic levitation, but a different mechanism, can explain the phenomenon at hand. This is further demonstrated in Sec. \ref{sec:Simulations}.

The magnetic interactions and external B-fields of static, uniformly magnetised spheres are precisely those of point dipoles at their centers\cite{edwards_interactions_2017,griffiths_introduction_2013}.
Thus in a dipole magnetic field, the floater magnet experiences the same force and torque as a point dipole of magnetic moment $\mathbf{m}_\text{f}$ at its center would.
In the quasistatic limit\cite{larsson_electromagnetics_2007} Faradays law of induction leads to the electric field $\mathbf{E} = -\partial_t \mathbf{A}$ where $\mathbf{A}$ is the vector potential, but except for the fields of induced currents, the magnetic fields are unchanged.
Thus using the well-known formulas for magnetic point dipoles\cite{griffiths_introduction_2013}, the fields from the rotor are
\begin{align}
    &\mathbf{B}_\text{r} = \frac{\mu_0}{4\pi r'^3} \left[3(\mathbf{\hat{r}'} \boldsymbol{\cdot} \mathbf{m}_\text{r}) \mathbf{\hat{r}'} - \mathbf{m}_\text{r}\right]
    \notag\\
    &\mathbf{E}_\text{r} = \frac{\mu_0}{4\pi} \frac{\mathbf{\hat{r}'} \times \mathbf{\dot{m}_\text{r}}}{r'^2} \label{eq:quasistatic_EM_fields}
\end{align}
where $\mathbf{m}_\text{r}$ is the magnetic moment of the rotor, $\mathbf{r'}$ is the position relative to the rotor center,
%, $\mathbf{\hat{r}'} = \mathbf{r'}/r'$
 and $\mu_0$ is the vacuum permeability. Also, $\mathbf{r'} = \mathbf{r} - \mathbf{d}$, where $\mathbf{r}$ is position relative to floater center, and $\mathbf{d}$ is the displacement vector from floater to rotor. We use\: $\hat{}$\: to denote normalised vectors and \:$\dot{}$\: for time derivatives.

The magnetic field leads to a dipole torque on the floater\cite{edwards_interactions_2017} $\boldsymbol{\tau}_\text{dip} = \mathbf{m}_\text{f} \times \mathbf{B}_\text{r}(\mathbf{r}=0)$, which on its own would make $\mathbf{m}_\text{r}$ and $\mathbf{m}_\text{f}$ anti-align, as this is the magnetostatic energy minimum when the positions are fixed. Anti-alignment is in direct contradiction with observation, so to explain the phenomenon, we require an opposing torque or dynamical effect to stabilise the observed moment configuration.

Clearly superconduction and feedback stabilisation are absent. The only non-electromagnetic couplings between the floater and the surrounding world are gravity, which does not exert a torque, and fluid effects such as centrifugal suction and air resistance, which either cancels it's own torque out or exerts a drag torque on the floater that is opposite to $\boldsymbol{\omega}_\text{f}$ \cite{schuck_ultrafast_2018, Lukerchenko2012, Turkyilmazoglu2022}.
This leaves electrodynamic effects, of which eddy current drag is the most significant, and gyroscopic effects. We discuss these in Secs.\ \ref{subsec:Electrodynamics} and \ref{subsec:Gyroscopic_stability}. It is well-established that dissipative and gyroscopic torques can induce steady-state configurations which are energetically unfavourable \cite{Krechetnikov_dissipation-induced_2007,Bou-Rabee_Tippe_2004}, and this is indeed what we observe numerically, as will be discussed subsequently in Sec. \ref{sec:Simulations}.

A hypothesis proposed by H. Ucar\cite{ucar_polarity_2021} is that due to a slight out-of-plane rotation of the rotor magnetization $\mathbf{m}_\text{r}$, or a shift of the rotor magnet relative to the rotation axis of its holder, $\mathbf{B}_\text{r}$ always has a vertical component, say $B_{\text{r}, z}$, at the floater center.
%, $\mathbf{r}=0$.
%This is opposed to perfectly symmetric alignment of rotor/floater in Fig. \ref{Fig.Angles} ensuring $\mathbf{B}_\text{r}(\mathbf{r}=0) = \mathbf{B}_{\text{r}, \perp}$ is entirely horizontal.
Since $B_{\text{r}, z}$ is constant, while the horizontal field $\mathbf{B}_{\text{r}, \perp}$ rotates rapidly, the former may have a greater impact on the floater dynamics than the small value of $B_{\text{r}, z}/B_{\text{r}, \perp}$ suggests. H. Ucar explains vertical stability by a repulsive force between the horizontal components of $\mathbf{m}_\text{f}$ and $\mathbf{B}_\text{r}$ countering an attractive force between the vertical components.

We propose that $B_{\text{r}, z}$ also dynamically stabilises the observed moment configuration. In Sec. \ref{sec:Simulations} we validate this assumption by reproducing the steady-state levitation and a number of experimental trends, using only magnetostatic dipole interactions and viscous drag. The drag coefficient required to reach the levitating state from rest is orders of magnitude greater than what can be explained by air resistance, suggesting that eddy currents or other electrodynamic loss channels are important.

\subsection{Electrodynamics \label{subsec:Electrodynamics}}
In the supplementary material we discuss electrodynamic effects in detail \cite{Supplementary_material}. The main takeaway is that the only potentially significant couplings, in addition to those of magnetostatic interactions between rotor and floater, are between the magnetic moments and induced eddy currents.
%The other torques have prefactors of the type $(\omega R_\text{f} / c)^2$ where $c$ is the speed of light, so these are entirely negligible.
Building on the existing literature\cite{hertz_induction_1880,reichert_complete_2012,mcdonald_conducting_nodate,youngquist_slowly_2016,nurge_thick-walled_2017,nurge_drag_2018,backus_external_1956,lorrain_electrostatic_1990,lorrain_magnetic_1998,yu_electromagnetic_2021,yu_optimal_2022} we derive (cf. the supplementary material \cite{Supplementary_material}) an analytical model for the coupling between the rotor moment and eddy currents in the conductive floater and find the torque from eddy currents to be
\begin{align}
    \boldsymbol{\tau}_\text{eddy} = \frac{2\pi}{15} \sigma (\omega_\text{r} - \omega_\text{f}) B^2_0 R^5_\text{f} \mathbf{\hat{z}}, \label{eq:tau_eddy}
\end{align}
Here, $\sigma$ is the conductivity, $R_\text{f}$ the floater radius, and
\begin{align*}
    B_0 = \frac{\mu_0}{4\pi d^3} m_\text{r}
\end{align*}
In deriving Eq. \eqref{eq:tau_eddy} we assumed the rotor is directly above the floater ($\mathbf{d}=d\mathbf{\hat{z}}$) and that both magnets only rotate around vertical ($\boldsymbol{\hat{\omega}}_\text{r} = \boldsymbol{\hat{\omega}}_\text{f} = \mathbf{\hat{z}}$), which does not capture the full range of motion. Also, we expand the rotor fields Eq. \ref{eq:quasistatic_EM_fields} in $R_f/d$ and only consider the leading order torque, which is equivalent to approximating $B_r$ as uniform. This is not justified, so Eq. \ref{eq:tau_eddy} should not be regarded as a general solution. That said, it should yield reasonable order-of-magnitude estimates and scaling relations.

If the rotor magnet is conductive, then the floater moment exerts a torque of the same form on the rotor. By angular momentum conservation, this leads to an equal and opposite torque on the floater. Thus eddy current coupling of the form in Eq. \eqref{eq:tau_eddy} is present when \textit{either} floater or rotor is magnetic and the other is conductive.
Additionally as the spinning floater induces eddy currents in the environment, it experiences further damping, which can easily exceed air resistance.
We observed experimentally that increasing this damping using an aluminium block qualitatively changes the dynamics. We note that in general coupling of rotor moment precession to floater rotation, as in Eq. \ref{eq:tau_eddy}, is distinct from the coupling of floater moment precession to rotor rotation. However, this subtlety is irrelevant for the subsequent argument.
For the state of motion considered here, the precession of the moments and the rotation of the bulk magnets is at the same frequency, while this is not necessarily the case in our simulations cf. Sec. \ref{subsec:Results_fixed_floater_position}.

%The conclusion is that eddy current torques are typically, significantly greater than air resistance, and tend to induce frequency locking but not quickly enough to explain experimental observations.

In the absence of all other torques, the equation of motion is $I_\text{f}\boldsymbol{\dot{\omega}}_\text{f} = \boldsymbol{\tau}_\text{eddy}$, where $I_\text{f}$ is the floater moment of inertia, in which case the difference in rotation frequency of the floater decays exponentially to the precession frequency of the rotor. This is the phenomenon of frequency locking by eddy current coupling. Using $I_\text{f}=\frac{8}{15}\pi \rho_\text{f} R_\text{f}^5$ where $\rho_\text{f}$ is the mass density of the floater, the exponential time constant can be written as $t_\text{eddy} = 4\rho_\text{f} / (\sigma B_0^2)$. At a relatively small levitation distance of $d = 20\:\mathrm{mm}$, and the experimentally relevant parameters $\mu_{0}M_\text{r} = 1.2\:\mathrm{T}$, $R_\text{r} = 9.5\:\mathrm{mm}$, $\rho_\text{f} = 7.5\:\mathrm{g/cm^3}$ and $\sigma = 667\:\mathrm{\Omega^{-1} mm^{-1} }$ we find $t_\text{eddy} = 25 \: \mathrm{s}$, while the experimentally observed frequency locking happens on fractions of a second. For comparison, if we consider only the dipolar torque, then $I_\text{f}\boldsymbol{\dot{\omega}}_\text{f} = \boldsymbol{\tau}_\text{dip} = \mathbf{m}_\text{f} \times \mathbf{B}_0$ and by dimensional analysis, we find the characteristic time $t_\text{dip} = \sqrt{I_\text{f}/(m_\text{f}B_0})$ which for the same parameter values and $R_\text{f} = 6.35\:\mathrm{mm}$ yields 1.7 ms. Thus, the frequency locking in our experiments must be driven by magnetostatic torques rather than eddy currents. We support this hypothesis in Sec. \ref{sec:Simulations} by numerical experiments.

In deriving Eq. (\ref{eq:tau_eddy}) we made several simplifying assumptions on the floater motion, so in reality $\boldsymbol{\tau}_\text{eddy}$ may have a horizontal component. This begs the question if $\boldsymbol{\tau}_\text{eddy}$ might stabilise the moment orientation. Since $\mathbf{m}_\text{f}$ is nearly vertical, $\tau_\text{dip} \approx m_\text{f}B_0$, so in our experiments
\begin{align*}
    \frac{\tau_\text{eddy}}{\tau_\text{dip}} \approx \frac{1}{30}\sigma(\omega_\text{r} - \omega_\text{f})R_\text{f}^2 \left(\frac{R_\text{f}}{d}\right)^3 \frac{m_\text{r}}{m_\text{f}} \leq 0.004
\end{align*}
This suggests eddy current effects are too small to balance the magnetostatic torque.
If $R_\text{f}$ was an order of magnitude larger, $\tau_\text{eddy}$ could be significant, but then self-induction and finite skin depth would also be important, necessitating a more complete analysis.

In summary, no electrodynamic effects can explain the moment orientation on their own.
%For larger magnets, eddy current coupling would produce frequency locking, but the effect is too slow at experimental parameters to match observation.
While including the non-uniformity of the rotor field would yield a more correct eddy current model, we doubt it would change this conclusion. Instead we propose that the moment configuration is produced by a vertical field component from imperfections in rotor placement, in conjunction with gyroscopic stabilisation.
%This we consider using a numerical model in the following.

\subsection{Gyroscopic stability \label{subsec:Gyroscopic_stability}}

For a floater with an anisotropic mass distribution, $I_\text{f}$ would be a 3-by-3 tensor and there would be an effective gyroscopic torque of the form\cite{fowles_analytical_2005} $\boldsymbol{\tau}_\text{gyro} = -\boldsymbol{\omega}_\text{f} \times  [I_\text{f}\boldsymbol{\omega}_\text{f}]$. This complication is absent in our experiments since we use spheres of uniform density.
One thing that does break the spherical symmetry is the intrinsic, angular momentum associated with the floaters magnetic moment. This gyromagnetic effect has been shown in theory to enable the stable levitation of a nanomagnet, even with a static applied field\cite{rusconi_quantum_2017,kustura_stability_2022}, however as shown in the supplementary material, the effect is negligibly small for the experiments described in this work \cite{Supplementary_material}.

Thus in the laboratory frame, no gyroscopic terms appear in the rotational equation of motion.
%, i.e.\ $I_\text{f} \boldsymbol{\dot{\omega}}_\text{f} = \boldsymbol{\tau}$ where $\boldsymbol{\tau}$ is the net torque from drag and electromagnetism.
That is not to say gyroscopic stability is absent, in fact in the rotor magnets' rest frame where its magnetic field is constant, there is the fictitious torque $\boldsymbol{\tau}_\text{fic} = -I_\text{f} \boldsymbol{\omega}_\text{r} \times  \boldsymbol{\omega}_\text{f}$, which bears a strong resemblance to $\boldsymbol{\tau}_\text{gyro}$. The cancellation of such a gyroscopic torque with gravity explains why a precessing top remains upright only when rapidly spinning \cite{fowles_analytical_2005}. Equivalently, in the laboratory frame, a certain angular acceleration is required to keep the top rotating, and gravity supplies this acceleration instead of making the top fall over.  Similarly, for the Levitron an initial spin around its magnetic moment is required to circumvent Earnshaws theorem and enable stable levitation\cite{berry_levitron_1996,simon_spin_1997,jones_simple_1997}.  In both cases, the argument also works for a spinning sphere.

For our experiments, there is no initial rotation, unlike the Levitron. Nevertheless, in our simulations the steady-state rotation during levitation is a combination of precession around vertical and spinning around the magnetic moment, in complete analogy to the spinning top. This state of rotation is gyroscopically stabilised against changes in the polar angle $\theta_\text{f}$. So in a sense, the rotation itself counteracts the magnetostatic torque. This is further discussed in Section \ref{subsec:Results_fixed_floater_position}.

\section{Simulations \label{sec:Simulations}}

\subsection{Model \label{subsec:Model}}
The rotors motion is fully constrained to a predefined trajectory. We model the corresponding time-evolution of the floater magnet using the following differential system
\begin{align}
    I\dot{\boldsymbol{\omega}}_\text{f} &= \mathbf{m}_\text{f} \times \mathbf{B}_\text{r} - \zeta_\text{rot} \boldsymbol{\omega}_\text{f} \label{eq:eqs_of_motion_rotation}
    \\
    \mathfrak{m} \dot{\mathbf{v}}_\text{f} &= \mathbf{F}_\text{dip} - \zeta_\text{trans} \mathbf{v}_\text{f} - g\hat{\mathbf{z}} \label{eq:eqs_of_motion_translation}
\end{align}
where $g=9.82\: \mathrm{m/s^2}$ is gravitational acceleration, $\mathbf{v}_\text{f}$ is floater velocity, $\mathbf{B}_\text{r}$ is the rotor field from Eq. \eqref{eq:quasistatic_EM_fields} evaluated at the floater's center, and
\begin{align}
    \mathbf{F}^\text{dip}_i &= \frac{3\mu_0}{4\pi} \frac{1}{r_\text{rf}^4}[(\mathbf{m}_\text{f} \mathbf{\cdot} \hat{\mathbf{r}}_\text{rf}) \mathbf{m}_\text{r} + (\mathbf{m}_\text{r} \mathbf{\cdot} \hat{\mathbf{r}}_\text{rf}) \mathbf{m}_\text{f}
    \notag\\
    &\quad +(\mathbf{m}_\text{f} \mathbf{\cdot} \mathbf{m}_\text{r})\hat{\mathbf{r}}_\text{rf}  -5(\mathbf{m}_\text{f}\mathbf{\cdot} \hat{\mathbf{r}}_\text{rf})(\mathbf{m}_\text{r} \mathbf{\cdot} \hat{\mathbf{r}}_\text{rf}) \hat{\mathbf{r}}_\text{rf}] \label{eq:F_dip}
\end{align}
is the dipole force from rotor on floater, where $\mathbf{r}_\text{rf}$ is the displacement from rotor to floater. We note that these magnetostatic interactions are exact for uniformly magnetised spheres\cite{edwards_interactions_2017}. For the drag coefficients, we use analytical solutions for viscous drag but with an effective viscosity to model eddy current damping, i.e.\
\begin{align*}
    \zeta_\text{rot} = 8\pi \eta_\text{eff} R_\text{f}^3 \quad , \quad \zeta_\text{trans} = 6\pi \eta_\text{eff} R_\text{f}
\end{align*}
This drag is exact for an isolated sphere at low Reynolds number\cite{rubinow_transverse_1961} and has the same linear dependence on angular velocity as we expect from eddy currents, confer Eq. \eqref{eq:tau_eddy}.
Unless otherwise stated, we used an efficient viscosity of $\eta = 1 \mathrm{Pa \cdot s}$ in the simulations. For reference, the viscosity of air at ambient conditions is $\eta_\text{air} = 0.02 \: \mathrm{mPa \cdot s}$, while water is $\eta_\mathrm{H_2O} = 1\: \mathrm{mPa \cdot s}$.

This type of model is well-established for magnetic nanoparticles, i.e.\ nanoscale, magnetised spheres suspended in liquid\cite{durhuus_simulated_2021}. That said, eddy current damping depends strongly on floater position, yielding more complex instabilities and transient behaviour, so we can at best reproduce qualitative trends and observations with the present model. The fact that the model does produce levitation suggests the phenomenon could also be achieved for entirely non-conducting magnets by other means than eddy current damping. The increased drag is possible by submerging the floater in viscous liquid.

One feature specific to the present model is in the angular velocity around the magnetic moment, i.e.\ $\omega_\text{m} = \mathbf{\hat{m}}_\text{f} \cdot \boldsymbol{\omega}_\text{f}$.
From Eq. \eqref{eq:eqs_of_motion_rotation} we have
\begin{align*}
    \dot{\omega}_\text{m} &= [\boldsymbol{\omega}_\text{f} \times \mathbf{\hat{m}}_\text{f}] \cdot \boldsymbol{\omega}_\text{f} + \frac{1}{I_\text{f}} \mathbf{\hat{m}}_\text{f} \cdot [\mathbf{m}_\text{f} \times \mathbf{B}_\text{r} - \zeta_\text{rot} \boldsymbol{\omega}_\text{f}]
    \\
    &= - \frac{\zeta_\text{rot}}{I_\text{f}} \omega_\text{m}
\end{align*}
That is $\omega_\text{m}$ can only decrease exponentially, so when it starts at zero, $\omega_\text{m} = 0$ at all times. We verified this numerically for all simulations. This is not in general true with eddy current coupling.

To simulate Eqs. \eqref{eq:eqs_of_motion_rotation} and \eqref{eq:eqs_of_motion_translation} we use velocity-Verlet timestep integration with a timestep of $1\: \mathrm{\mu s}$\cite{young2014leapfrog}. To rotate $\mathbf{m}_\text{f}$ we use the Euler-Rodrigues' rotation formula\cite{cheng_historical_1989}.

\subsection{Results: fixed floater position \label{subsec:Results_fixed_floater_position}}

\renewcommand{\thefigure}{6}
\renewcommand{\figurename}{Fig.}
\begin{figure*}[t]
    \includegraphics[width=\textwidth]{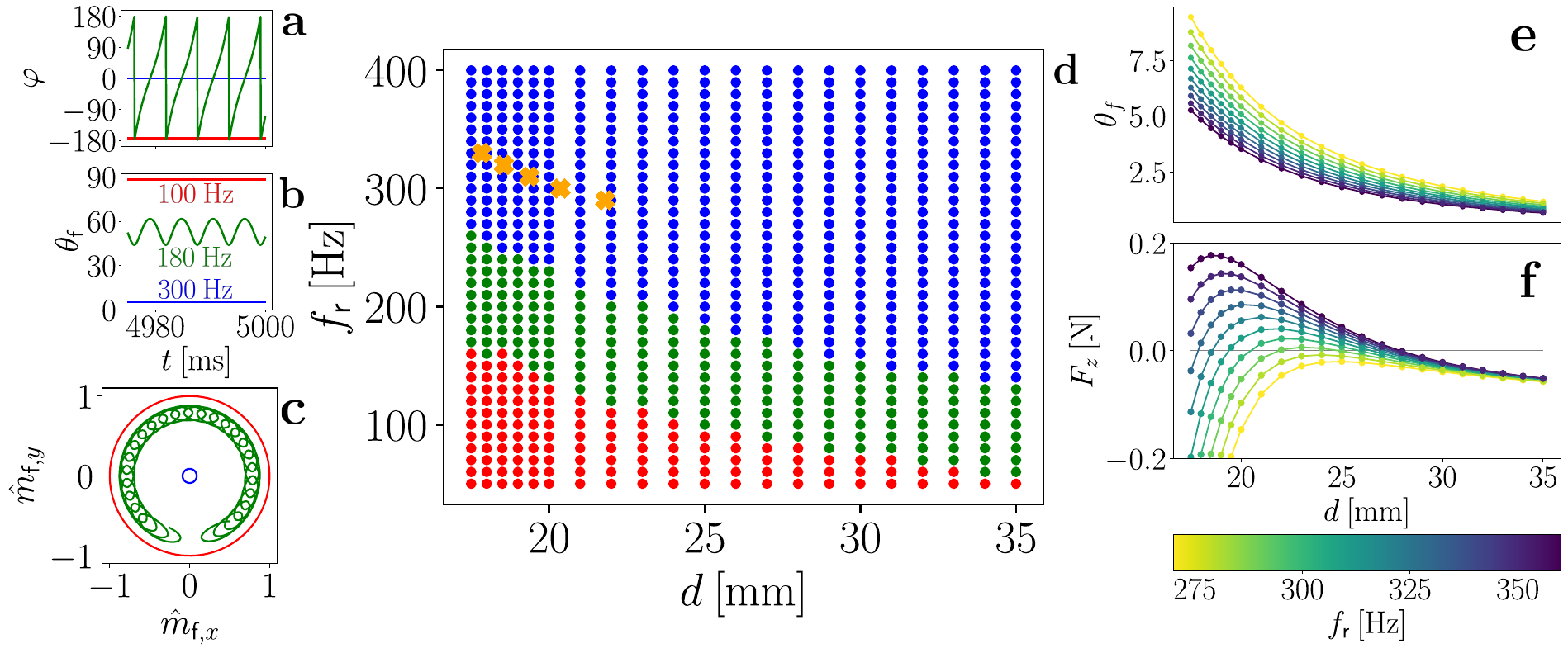}
    \caption{Simulation results at various rotor frequencies with the floater position fixed. The rotor is displaced by $\delta_\text{r} = 1\:\mathrm{mm}$ relative to the rotation axis, yielding a slight vertical B-field, the effective damping is $\eta = 1\:\mathrm{Pa\cdot s}$, both magnetisations are 1.18 T and the diameters are 12.7 mm and 19 mm for floater and rotor respectively. \textbf{a-c)} Steady-state behaviour in the 3 different dynamical phases for $d=20\:\mathrm{mm}$. \textbf{a-b)} phase lag, $\varphi$, and floater polar angle, $\theta_\text{f}$, vs. time over 25 ms. \textbf{c)} Transverse components of the normalised floater moment over 150 ms. \textbf{d)} The dynamical phase of every simulation at various rotor frequencies, $f_\text{r}$, and rotor/floater distances $d$. Orange crosses mark stable force equilibria. \textbf{e-f)} $\theta_\text{f}$ and vertical force $F_z$ for all simulations in the blue phase between 270 and 360 Hz. The stable equilibria are the points where spline-fits of the force curves in fig.\ \textbf{f)} cross from positive (attractive) to negative (repulsive) as $d$ decreases.}
    \label{Fig.freq_sweep}
\end{figure*}

We find that for most parameters, the simulated floater will either fall under gravity or collide with the rotor and be flung away. Therefore as a first study, we keep the center position of the floater fixed and only consider the rotational motion.

For all simulations, the floater is initially at rest and centered on the rotation axis, which is vertical. We find that the steady-state is not sensitive to initial orientation, so we only used $60^\circ$ from vertical in the $xz$-plane, i.e.\ $\hat{\mathbf{m}}_\text{f}(t=0) = (\sin(60^\circ), 0^\circ, \cos(60^\circ))$. The rotor configuration is given in general by
\begin{align}
    \mathbf{r}_\text{r} = \begin{pmatrix}
    \delta_\text{r} \cos(2\pi f_\text{r} t) \\ -\delta_\text{r} \sin(2\pi f_\text{r} t) \\ d
    \end{pmatrix}
    \quad , \quad
    \mathbf{m}_\text{r} =
    \begin{pmatrix}
        \cos(\varphi_0 - 2\pi f_\text{r}t) \\ \sin(\varphi_0 - 2\pi f_\text{r}t) \\ 0
    \end{pmatrix}   \label{eq:rotor_trajectory}
\end{align}
That is, the rotor magnet is a distance $d$ vertically above the floater, except for a small lateral shift of $\delta_\text{r}$ away from the rotation axis. Its holder rotates clockwise at frequency $f_\text{r}$ with the moment in the $xy$-plane and pointing away from the rotation axis. We used $\varphi_0 = 60^\circ$ as initial condition. We distinguish between distance $r_\text{rf}$ and vertical distance $d$, but since $\delta_\text{r} \ll d$, they are nearly equal. Besides the lateral shift, the setup is identical to Fig. \ref{Fig.Angles}.

With the floater centered on the rotation axis, the only effect of $\delta_\text{r}$ is to produce a constant, vertical B-field. Since $\mathbf{m}_\text{r}$ points away from the rotation axis, $B_{\text{r},z} > 0$, while if $\mathbf{m}_\text{r}$ pointed towards the axis we would have $B_{\text{r},z} < 0$. The same effect can be achieved by giving $\mathbf{m}_\text{r}$ an out-of-plane tilt. We verified that for either sign $m_{\text{f},z}$ points in the field direction.

Without a vertical field, the steady-state always has $\mathbf{m}_\text{f}$ in the $xy$-plane, corresponding to a polar angle of $\theta_\text{f} = 90^\circ$. While $\delta_\text{r} = 1 \: \mathrm{mm}$ seems small, for some frequencies the resulting $B_{\text{r},z}$ is sufficient to make $\mathbf{m}_\text{f}$ nearly vertical. We find that depending on the system parameters, there are 3 different dynamical phases with radically different steady-state behaviour and sharp transitions in parameter space.

In Fig. \ref{Fig.freq_sweep} we present simulation data for a range of different frequencies and distances. Figs. \ref{Fig.freq_sweep}a-c illustrate the 3 ("red", "green", "blue") dynamical phases. Fig. \ref{Fig.freq_sweep}a shows that at low and high $f_\text{r}$ the phase angle is essentially constant with a value close to either $180^\circ$ or $0^\circ$ (red and blue curves, respectively), i.e.\ rotor and floater are frequency locked. At intermediate frequencies, $\mathbf{m}_\text{f}$ rotates more slowly than $\mathbf{m}_\text{r}$, but there is a secondary precession at equal frequency, hence the fast oscillations on the green curves in Figs. \ref{Fig.freq_sweep}b-c. We note that when increasing $f_\text{r}$ or $d$ the time averaged value of $\theta_\text{f}$ decreases gradually from $90^\circ$ to $0^\circ$.

In Fig. \ref{Fig.freq_sweep}d we see the distribution of dynamical phases in parameter space. Each data point corresponds to a 5 s simulation, where we computed average and standard deviation of $\varphi$ over the last 250 ms. Green points have standard deviations above $1^\circ$, while blue and red have below $1^\circ$. Except for a few points on the phase boundaries, the standard deviation is either above $50^\circ$, or less than $10^{-6}$, i.e. there is a sharp contrast, so the phase determination for a given simulation is unambiguous. To distinguish blue and red, we use the steady-state value of $\varphi$. In the supplementary material, we show similar plots where we varied $\eta$, $R_\text{f}$, $M_\text{f}$ or $\delta_\text{r}$ instead of $f_\text{r}$ \cite{Supplementary_material}.

The red phase is the energy minimum which would be the final configuration with a static rotor, i.e.\ when $f_\text{r} = 0$. Thus it makes sense the red phase occurs at lowest frequencies, and at shorter distances where the energy minimum is deepest. At intermediate frequencies and distances, inertia and drag means the floater cannot keep up, so while there is a net clockwise rotation, it is only the smaller, secondary precession which follows the rotor frequency. As the frequency increases, the floater is less and less susceptible to the rotating, horizontal field, but still responds to the constant, vertical field, hence $\mathbf{m}_\text{f}$ becomes increasingly vertical. When $\theta_\text{f}$ is near 0, the secondary precession disappears and frequency locking is reestablished, i.e.\ the green phase transitions to blue.

To explain why the blue phase occurs, we first infer an expression for the angular velocity. We know that $\dot{\varphi} = 0$, so there is a clockwise precession around vertical at a frequency of $\omega_\text{r}$. Also $\dot{\theta}_\text{f} = 0$, so the only possible form of rotation is spinning around $\mathbf{m}_\text{f}$. That is
$\boldsymbol{\omega}_\text{f} = -\omega_\text{r} \mathbf{\hat{z}} + \dot{\psi} \mathbf{\hat{m}}$. Using that $\mathbf{m}_\text{f} \cdot \boldsymbol{\omega}_\text{f} = 0$, as discussed in Section \ref{subsec:Model}, we find that $\dot{\psi} = \omega_\text{r} \cos \theta_\text{f}$, hence the angular velocity is
\begin{align}
    \boldsymbol{\omega}_\text{f} = \omega_\text{r} [-\mathbf{\hat{z}} + \cos \theta_\text{f} \mathbf{\hat{m}}]  \quad \text{(Blue phase)}
    \label{eq:omega_blue_phase}
\end{align}
We verified that each component fits the simulations to within a relative error of $0.0004\:\%$ throughout the blue phase.
%This is analogous to \cite[eq. 17]{ucar_polarity_2021}, however we disagree on the exact form.
The magnitude is simply $\omega_\text{f} = \omega_\text{r} \sin \theta_\text{f}$.

Counter-intuitively, the floater spins around its moment precisely because the \textit{net} angular velocity around $\mathbf{m}_\text{f}$ must be 0 and there is a contribution from the precession of $\mathbf{m}_\text{f}$ around vertical.
%This state of rotation, i.e.\ spinning around an axis that itself precesses around vertical, is the same as a spinning top under gravity.
This spinning gyroscopically stabilises against changes in $\theta_\text{f}$, analogously to how a spinning top defies gravity\cite{fowles_analytical_2005}. Equivalently, in the rest-frame of the rotor the spinning yields a fictitious, gyroscopic torque that balances the magnetostatic torque. When $\mathbf{m}_\text{f}$ is close to vertical, the angular velocity vectors from precession and from spinning around $\mathbf{m}_\text{f}$ nearly cancel, so that the magnitude of $\omega_\text{f}$ is tiny. Consequently the drag torque is small and does not hinder frequency locking.

By inserting Eq. \ref{eq:omega_blue_phase} in the governing equation Eq. \ref{eq:eqs_of_motion_rotation} we get a set of equations $\theta_\text{f}$ and $\varphi$ must obey for self-consistency. In the supplementary material we expand said equations to leading order in the small angles $\theta_\text{f}, \varphi$ and isolate, which yields \cite{Supplementary_material}
\begin{align}
    \varphi = \frac{\zeta_\text{rot} \omega_\text{r}}{I_\text{f}\omega_\text{r}^2 - m_\text{f}B_{\text{r}, z}} \quad \text{and} \quad \theta_\text{f} = \frac{m_\text{f}B_{\text{r}, \perp}}{I_\text{f}\omega_\text{r}^2 - m_\text{f}B_{\text{r}, z}}, \label{eq:angles_blue_phase}
\end{align}
Comparing to the blue simulations in Fig. \ref{Fig.freq_sweep}d, the greatest relative error is $1.8\%$ for $\varphi$ and $0.05\%$ for $\theta_\text{f}$, so higher order corrections are more significant for $\varphi$. From Eq. \ref{eq:angles_blue_phase} we see that a finite moment of inertia is required for the blue phase to occur, because a negative value of $\theta_\text{f}$ is unphysical, but drag is not required for self-consistency. $I_\text{f} \sim R_\text{f}^5$, so the importance of inertia may explain why the critical frequency has a strong size dependence despite a weak dependence on magnetisation, as seen in Fig. \ref{Fig.LevDistSizeMag}a and the supplementary material \cite{Supplementary_material}.

The next question is how the computed rotational states can produce stable levitation. In the red phase, the magnets are consistently attractive. In the green phase, the force oscillates at a frequency of $f_\text{r}$, but it averages to a net repulsion; at least for all parameter combinations we simulated. While this repulsion could balance gravity in special cases, our experiments showed stability regardless of whether the floater was above or below the rotor, so this would be a different form of levitation. That leaves the blue phase where $\varphi$, $\theta_\text{f}$ and the vertical force $F_z$ are all constant.

In Fig. \ref{Fig.freq_sweep}f we show the vertical force for frequencies from 270 to 360 Hz. It is apparent that the force curves are non-monotonous. This is because the magnetic force changes from attractive ($F_z > 0$) to repulsive ($F_z < 0$) with decreasing distance. Every point where $F_z$ crosses 0 from positive to negative as $d$ decreases is a stable equilibrium, because a slight displacement in either direction leads to a restoring force. In some cases the maximum attractive force is less than gravity, in others the magnets would collide before reaching the stable point. In Fig. \ref{Fig.freq_sweep} we marked all mid-air, stable points by orange crosses. We note that the levitation distance decreases with increasing frequency, in agreement with Fig. \ref{Fig.ModesQuantified}.

To understand the force curves, we note that the blue phase is close to the configuration $\theta_\text{f} = \varphi = 0^\circ$ and $\delta_\text{r}/d \ll 1$. Therefore, we consider a Taylor expansion of the dipole force, Eq.\ \eqref{eq:F_dip}, in these 3 variables.
To zeroth order, $\hat{\mathbf{m}}_\text{f} = \hat{\mathbf{r}}_\text{rf} = \hat{\mathbf{z}}$ which is perpendicular to $\mathbf{m}_\text{r}$, in which case the vertical force is 0.
To first order
\begin{align*}
    F_{\text{dip},z} \approx \frac{3\mu_0 m_\text{r} m_\text{f}}{4\pi d^4} \left[ 4 \frac{\delta_\text{r}}{d} - \theta_\text{f}\right] \quad \text{Blue phase}
\end{align*}
The first term is an attractive force between the vertical components of $\mathbf{B}_\text{r}$ and $\mathbf{m}_\text{f}$. The second term is a repulsion between horizontal components. From Fig. \ref{Fig.freq_sweep}e we see that $\theta_\text{f}$ increases with decreasing distance, and the relative change is greater than that in $\delta_\text{r}/d$. This is why $F_{\text{dip},z}$ can change from attractive to repulsive in mid air. In other words, as the magnets approach, the floater moment becomes more horizontal, which makes the repulsive dipole force grow \textit{relative to} the attractive dipole force.
This explanation is in agreement with Ref. \cite{ucar_polarity_2021}.
\subsection{Results: free floater position \label{subsec:Results_all_DoF}}

\renewcommand{\thefigure}{10}
\renewcommand{\figurename}{Video}
\begin{figure*}[t]
    \includegraphics[width=\textwidth]{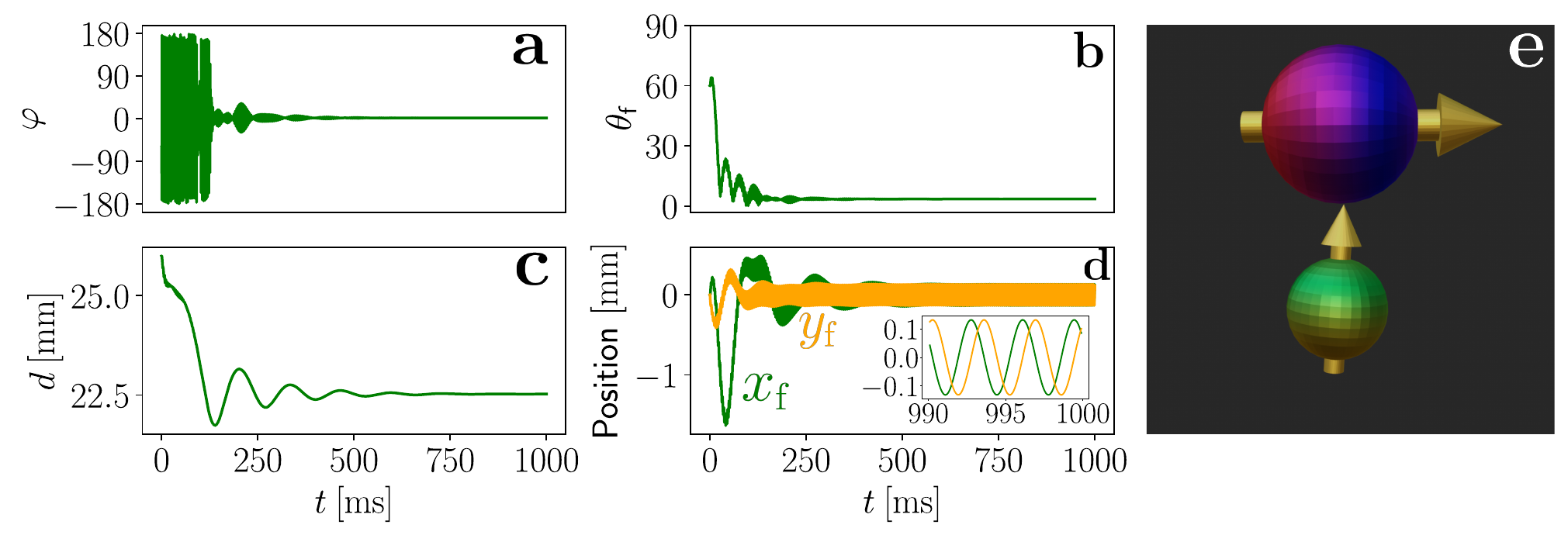}
    \caption{Simulated levitation at a rotor frequency of $f_\text{r} = 300 \: \mathrm{Hz}$ where the floater has full rotational and translational freedom. Same parameters as in Fig. \ref{Fig.freq_sweep} except $\delta_\text{r} = 1.2\: \mathrm{mm}$. \textbf{a)} phase angle. \textbf{b)} floater polar angle. \textbf{c)} vertical floater/rotor distance. \textbf{d)} xy-position of floater with origin on the rotors rotation axis. The inset in \textbf{d)} shows the last 10 ms of the 1 s simulation. \textbf{e)} final configuration. Blue/red sphere is rotor, green/orange is floater and yellow arrows represent magnetic moments. Positions and sizes are to-scale. Direct link: \href{https://youtu.be/875UmmfUcYE}{Video 10.}}
    \label{Vid.levitation_sim}
\end{figure*}

Here, the rotor is fully constrained as in Eq. \eqref{eq:rotor_trajectory}, while the floater is free to both rotate and translate in 3-dimensional space. We find that with the right parameters, the model (Eqs. \eqref{eq:eqs_of_motion_rotation} and \eqref{eq:eqs_of_motion_translation}) does produce stable levitation. In Video. \ref{Vid.levitation_sim} is shown an example simulation at $f_\text{r} = 300 \: \mathrm{Hz}$ and the same parameters as Fig. \ref{Fig.freq_sweep} except $\delta_\text{r} = 1.2\:\mathrm{mm}$. We were unable to produce levitation at $\delta_\text{r} = 1\:\mathrm{mm}$, which we attribute to the translational motion making frequency locking less stable. For the same reason, the steady-state vertical distance of $d = 22.5 \: \mathrm{mm}$ is slightly higher than expected from Fig. \ref{Fig.freq_sweep}d, i.e. from the orange cross at 300 Hz.

We observe from Video \ref{Vid.levitation_sim}a-b that we get the expected dynamical phase and from Video \ref{Vid.levitation_sim}c-d that the levitation is stable in all directions. In steady-state, the floater $z$-coordinate and the relative orientation of the magnets settle to constants, while the floater's $xy$-coordinates perform sinusoidal oscillations, just like the Side mode observed experimentally (see inset in Video \ref{Vid.levitation_sim}d). Note how all configurational variables reach steady-state within half a second, which is much faster than frequency-locking by eddy-currents, cf. Section \ref{subsec:Electrodynamics}.

We got qualitatively identical results when generating the vertical B-field by giving $\mathbf{m}_\text{r}$ a $5^\circ$ out-of-plane tilt, rather than a lateral shift of the whole magnet. This is non-trivial as $\mathbf{B}_\text{r}$ scales differently with floater position for the two imperfections in rotor placement. Indeed H. Ucar found experimentally that levitation also occurs when the constant field is from a third, stationary dipole magnet \cite[Fig. 20]{ucar_polarity_2021}. We conclude that the key aspect which enables this novel form of magnetic levitation is the superposition of stationary and time-varying magnetic fields, not the specific rotor perturbations.

When letting the system relax at $\eta = 1\:\mathrm{Pa\cdot s}$ for 1 s then turning off all damping, the final configuration remains stable indefinitely. It still exhibits the Side mode. This shows that damping is not strictly required to sustain the moment orientation that yields levitation. Rather damping serves to facilitate reaching the levitating state from rest and to increase the stability thereof.

\subsection{Comparison of simulations and experiments}

In Sections \ref{subsec:Model}, \ref{subsec:Results_fixed_floater_position} and \ref{subsec:Results_all_DoF} we have shown that a model including only magnetostatic (dipole) coupling and damping can reproduce stable levitation in certain parameter ranges. The effect of gravity is to shift the point of stability, but it alters none of the qualitative features. The model has no features unique to eddy currents as the same damping can be achieved by putting the floater in liquid, and while moderate viscous damping facilitates levitation, it is not a strict requirement.

This suggests the levitation effect is quite general, and since dipole interactions are significant across many length-scales, it should be scalable. The relatively small moment of inertia might be an issue for the levitation of microparticles, but perhaps this can be compensated by increasing the frequency of the rotating field.

The simulations are qualitatively consistent with the experiments in terms of how levitation distance scales with rotor speed (compare Figs. \ref{Fig.ModesQuantified}b and \ref{Fig.freq_sweep}d) and with floater size (see Fig. \ref{Fig.LevDistSizeMag}b and supplementary material \cite{Supplementary_material}). These are strong indications that the model contains most of the essential physics.

Quantitatively, the simulations have room for improvement. From the orange crosses in Fig. \ref{Fig.freq_sweep}d we see that levitation is predicted at higher frequencies and a narrower frequency band than found experimentally (see Fig. \ref{Fig.ModesQuantified}b). Also while Video \ref{Vid.levitation_sim}d is consistent with the side mode in Video \ref{Video_side}, we have not observed any vertical oscillations in steady-state, so our simulations do not reproduce the Up-down, U-shaped or Mixed mode. Regarding the semi-stable mode, when our simulations reach steady-state, the system remains stable indefinitely, even without damping.
%do not numerically observe any growth in this mode, nor a change in the equilibrium distance, so our simulations remain stable indefinitely. We have not observed the up-down, U-shaped or mixed mode either, so our numerical model should primarily be regarded as an explanatory model of the steady-state behaviour.

Experimentally, we do observe indefinite stability when an aluminium block is placed below the floater, which has the effect of increasing damping. So our numerical model is perhaps more representative of this setup. While the point of stability shown in Video \ref{Vid.levitation_sim} is largely consistent with Fig. \ref{Fig.freq_sweep}d, the floater orientation and force curve
%distribution of dynamical phases
will change at least slightly when including floater translation. In addition the system is highly sensitive to the constant B-field, so quantitative experimental comparison requires a more accurate determination of the rotor magnets trajectory, i.e.\ $\delta_\text{r}$ and $\theta_\text{r}$. Alternatively, one can use a stationary, third magnet to generate the constant field.

On the modelling side, we believe the main point of improvement is the drag. At $f_\text{r} = 200\: \mathrm{Hz}$, we find that a viscosity of at least $0.01 \: \mathrm{Pa \cdot s}$ is required to produce the blue phase from rest (see supplementary material \cite{Supplementary_material}), which cannot be explained by air resistance.
%Also, even accounting for the large relative uncertainty the measured phase angle is at least $\varphi \approx 1^\circ$. Comparing with Eq.\ \ref{eq:angles_blue_phase} at $f_\text{r} = 200\: \mathrm{Hz}$ and noting that $I_\text{f} \omega_\text{f} \gg m_\text{f} B_{\text{r},z}$, we estimate a viscosity of $\eta = 0.5\:\mathrm{Pa\cdot s}$. This is a very rough estimate, but still a clear indication that air resistance is insufficient.
If, as we suspect, the main source of damping is eddy currents, then this has a number of implications. The relative magnitude of translational and rotational damping in Video \ref{Vid.levitation_sim} is most likely incorrect, Eq. \eqref{eq:tau_eddy} suggests a stronger dependence on floater size and magnetisation than the simulated model, and damping would be a function of position. Finally there is a distinction between eddy current \textit{damping} from the stationary environment, and eddy current \textit{coupling} between the two rotating magnets, which can also accelerate the floater.

A more accurate model of damping, rotational and translational, may be required to explain the relatively complex set of translational modes or the gradual change in equilibrium distance observed in the semi-stable mode. In particular we note that our explanation for the stability of the moment orientation relies heavily on the angular velocity component $\omega_\text{m} = \mathbf{\hat{m}} \cdot \boldsymbol{\omega}_\text{f}$ being constantly 0. We hypothesize that the breaking of this conservation law by eddy current coupling is related to the very gradual decrease in levitation height over time recorded in Fig. \ref{Fig.FallingRate}, but further investigation is required.

%One potential application is the trapping and contactless manipulation of ferromagnetic microparticles. This begs the question how well the phenomenon scales with floater size.
%The simulated model has no features unique to eddy currents as the same damping can be achieved by putting the floater in liquid, gravity is unnecessary, and while moderate viscous damping facilitates levitation, it is not a strict requirement.
%Dipole interactions are significant across many length scales, so the main limitation is that the moment of inertia scales with $R_\text{f}^5$.
%, which is part of the reason inertia is typically negligible at the microscale\cite{purcell_life_1977}.
%Perhaps the low inertia of microparticles can be compensated by higher field frequencies, but further investigation is required.

\section{Conclusion}
We have demonstrated that magnetic levitation of a permanent magnet can be achieved by placing it in the vicinity of another magnet rotating at angular velocities in the order of
$200 \: \mathrm{Hz}$.
%$10000\;\text{RPM}$.
Rotating a 19 mm diameter spherical NdFeB magnet, the levitation phenomenon was demonstrated for spherical magnets in the 5-30 mm diameter range. First, we showed experimentally that the floating magnet aligns to the like pole of the rotating magnet, i.e. oppositely to what would be expected from magnetostatics. For a floating magnet with a diameter of 12.7 mm we measured the polar and azimuthal (phase) angles as $\theta_f=7^\circ\pm4^\circ$ and  $\varphi=6.4^\circ\pm5.1^\circ$, respectively.
Then we investigated the instabilities occurring during levitation, and observed five different patterns of motion. Afterwards, we assessed the influence of rotation speed, as well as the size and magnetization of the floating magnet on levitation. We found that the floating magnet's magnetization does not affect the rotation speed needed for levitation (in case of remanence fields of 760 - 1180 mT), while its size has a very clear influence: the smaller the floating magnet, the higher the rotor speed necessary to achieve levitation and the further the floating magnet levitates from the rotating magnet.
%Subsequently, we discussed the underlying physics of the system. A thorough analysis revealed that electrodynamics is orders of magnitude too small to sustain levitation. However, using a numerical model we showed that a small, constant, vertical magnetic field component, which may originate from imperfections in the positioning of the rotor magnet, in conjunction with eddy current dissipation, reproduces the experimental trends observed.
Subsequently, we discussed the underlying physics of the system. A thorough analysis revealed that electrodynamic effects are orders of magnitude too small to sustain the levitating state. However, using a numerical model involving only magnetostatics and eddy-current enhanced damping, we reproduced experimental trends and stable levitation. The key is a constant, vertical magnetic field component, which may originate from small imperfections in the positioning of the rotor magnet. This enables the counter intuitive steady-state moment orientation, which is gyroscopically stabilised and in turn gives the magnetostatic force a stable, mid-air equilibrium point.

\section*{Data statement}
All data presented in this work are available from Ref. \cite{Data_2023}. This includes the videos of all experiments described in the text. The supplementary illustrative videos can also be found at Ref. \cite{YoutubeVideos}.

\section*{Acknowledgements}
We wish to thank Sintex a/s, Jyllandsvej 14, 9500 Hobro, Denmark, specifically Flemming Buus Bendixen, for providing the epoxy magnets used in parts of the experiments.

\bibliographystyle{unsrt}


\clearpage

%\includepdf[pages=-]{supplementary.pdf}
\begin{strip}
    \huge{Supplementary material for the article ``Magnetic levitation by rotation''}
\end{strip}
%\twocolumn[\centerline{\huge{Supplementary material for the article}}]
%\twocolumn[\centerline{\huge{``Magnetic levitation by rotation''}}]

\begin{strip}
\setcounter{section}{0}

\section{Determination of the phase angle between rotor and floater}
To determine the phase angle between the rotor and floater magnet, we analyse the slow-motion video of the painted magnets.  The edge of the paint on the left side is tracked to determine the orientation of the floater magnet. In the same frames the hue of the rotor magnet corresponding to its magnetization direction was also recorded. Each of these signals are sinusoidal and by taking the difference between them we can find the phase. This was done for two experiments that produced similar results, both of which are shown in Fig.  \ref{Fig.PhasePlot_2}. As can been from the figure the floater and the rotor magnet are nearly in phase with a small phase shift of $\varphi=6.4^{\circ}\pm5.1^{\circ}$.
%To determine the phase angle between rotor and floater from the tracked video, the following procedure was applied. In the video the y-position of the tracked point on the floater magnet that separates its north and south hemispheres was measured. In the same frames, the hue of the rotor magnet, corresponding to its magnetization direction, was also recorded. Both of these measurement are sinusoidal in nature, and shown in Fig. \ref{Fig.PhasePlot_2} is the difference between the two sinusoidal signals in arbitrary units, as function of the phase angle, $\varphi$, which is then also directly the phase angle shown in Fig. \ref{Fig.PhasePlot_2}. As can be seen, the signals are very close to being in phase. This was done for two experiments, which produced very identical results.

\renewcommand{\thefigure}{1}
\renewcommand{\figurename}{Figure Supp.}
%\begin{figure*}[!h]
\begin{center}
\includegraphics[width=1\columnwidth]{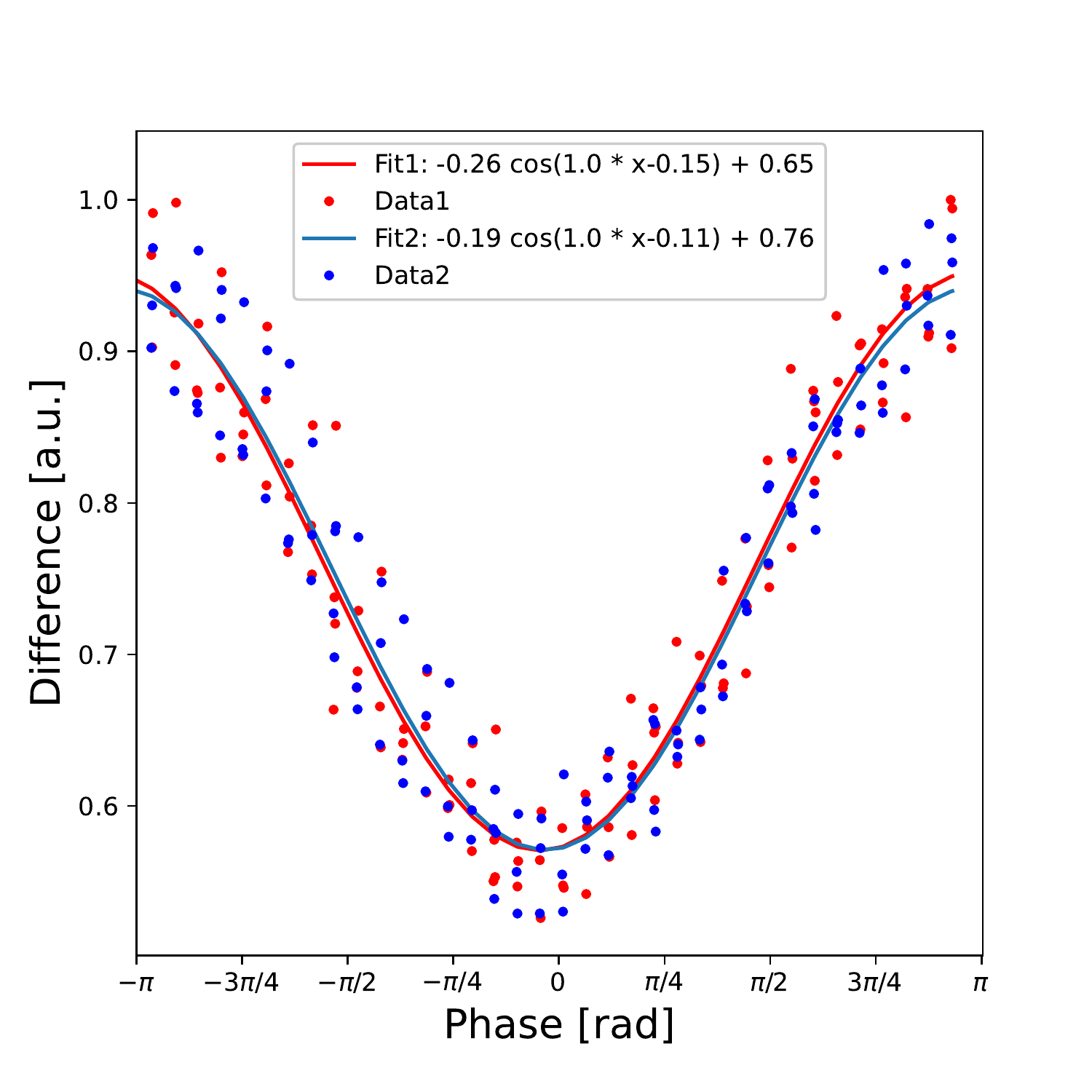}
\captionof{figure}{The phase between the floater and the rotor magnet. Calculated from synchronising the rotor magnet and the floater magnet's movement. The experiments were done at 200 Hz using the floater with a diameter of 12.7 mm.}
\label{Fig.PhasePlot_2}
\end{center}
%\end{figure*}

\section{Levitation distance in a high damping case.}
To test if the magnets can levitate forever an experiment was conducted using rotating a 12. 7mm magnet at 240 Hz keeping an aluminum plate 33.5 mm below the rotor magnet to dampen instabilities. It was found that the motion would initially fall with a rate of 0.3 mm/s and after 20 seconds it drastically slows down to a speed of around $1\: \frac{\mu\text{m}}{\text{s}}$. This slow speed decreases further until it finds an equilibrium height. The experiment is conducted for 40 minutes with no sign of stopping.

\renewcommand{\thefigure}{2}
\renewcommand{\figurename}{Figure Supp.}
%\begin{figure*}[!h]
\begin{center}
\includegraphics[width=1\columnwidth]{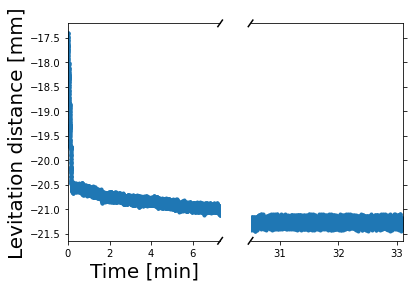}
\captionof{figure}{The levitation distance between the floater and the rotor magnet with an aluminum plate 33.5 mm away from the rotor magnet. The experiments were done at 240 Hz using a floater magnet with a diameter of 12.7 mm.}
%\caption{The levitation distance between the floater and the rotor magnet with an aluminum plate 33.5 mm away from the rotor magnet. The experiments were done at 240 Hz using a floater magnet with a diameter of 12.7 mm.}
\label{Fig.PhasePlot_3}
\end{center}
%\end{figure*}

\section{Effect of magnetic angular momentum}

It is well-known that changes in magnetisation can make an object rotate (Einstein--de-Haas effect), and that spinning an object can magnetise it (Barnett effect).
%\cite{barnett_magnetization_1915}.
Both effects stem from the angular momentum associated with magnetic moments, i.e.\ $\mathbf{J}_\text{mag} = -\gamma^{-1} \mathbf{m}$ where $\mathbf{m}$ is magnetic moment and $\gamma = 1.76\cdot 10^{11} \:\mathrm{Hz/T}$ is the gyromagnetic ratio. Below we show that $\mathbf{J}_\text{mag}$ is negligible for the levitating magnet phenomenon. All quantities are defined in the main text.

By the definition of angular velocity we have $\mathbf{\dot{m}}_\text{f} = \boldsymbol{\omega}_\text{f} \times \mathbf{m}_\text{f}$. We just consider the magnetostatic coupling, so the net torque on the floater is $\mathbf{m}_\text{f} \times \mathbf{B}_\text{f}$.
Using angular momentum conservation, the equation of motion for the floater is then
\begin{align*}
    I_\text{f} \boldsymbol{\dot{\omega}}_\text{f} - \gamma^{-1} \mathbf{\dot{m}}_\text{f} = \mathbf{m}_\text{f} \times \mathbf{B}_\text{r} \xRightarrow{} I_\text{f} \boldsymbol{\dot{\omega}}_\text{f} = \mathbf{m}_\text{f} \times \mathbf{B}_\text{r} + \gamma^{-1} \boldsymbol{\omega}_\text{f} \times \mathbf{m}_\text{f} = \mathbf{m}_\text{f} \times \left[\mathbf{B}_\text{r} - \gamma^{-1}\boldsymbol{\omega}_\text{f} \right]
\end{align*}
Thus including $\mathbf{J}_\text{mag}$ in the model results in an effective magnetic field of the form $-\gamma^{-1}\boldsymbol{\omega}_\text{f}$. At a frequency of 400 Hz, which is the highest we tested, the effective field has a magnitude of 14 nT.  This is clearly negligible.
%We note that including additional torques, e.g.\ from air resistance or electrodynamics, does not change the conclusion.

\section{Eddy current coupling \label{subsec:eddy_current_coupling}}

In this section we analyse electrodynamic couplings between the rotor and floater. In particular by approximating the rotor field as uniform across the floater, we solve analytically for the charge and current distributions in the floater. The main result is the eddy current coupling in Eq. \eqref{eq:tau_eddy}, i.e.\ the torque exerted by the rotor field on currents induced in the conductive floater. We show in Sec. \ref{subsec:Electrodynamic_torques} that all other electrodynamic couplings are negligible in the non-relativistic limit.

Because of Lorentz forces on the conduction electrons, the floater acquires an eddy current density at position $\mathbf{r}$ relative to its center of\cite{griffiths_introduction_2013,mcdonald_conducting_nodate}
\begin{align}
    \mathbf{J}_\text{eddy} = \sigma (\mathbf{E} + \mathbf{v} \times \mathbf{B}) \quad , \quad \mathbf{v} = \mathbf{v}_\text{CM} + \boldsymbol{\omega} \times \mathbf{r}, \label{eq:J_eddy}
\end{align}
where $\sigma$ is conductivity, $\mathbf{v}(\mathbf{r})$ is velocity of the floater at position $\mathbf{r}$ and $\mathbf{v}_\text{CM}$ is center-of-mass velocity. This is in addition to the convective current $\rho \mathbf{v}$ from the charge distribution $\rho$ moving along with the floaters mechanical motion.
% This is in addition to a Faraday charge distribution $\rho$ and a convective current $\rho \mathbf{v}$ from the Faraday charges following rotating along with the floater, however these effects.

%Eddy currents are well-known from Maglev trains and induction motors\cite{mcdonald_electrodynamics_nodate}. It has been demonstrated that due to forces and torques from $B$-fields on induced eddy currents, magnetic fields can be used to levitate and rapidly spin a conductive sphere\cite{schuck_ultrafast_2018}. Also spinning dipoles can be used for contactless manipulation of the position and orientation of metal spheres\cite{pham_dexterous_2021,tabor_adaptive_2022,dalton_attracting_2022}.

Eddy currents are well-known from Maglev trains and induction motors. It has been demonstrated that due to forces and torques from $B$-fields on induced eddy currents, magnetic fields can be used to levitate and rapidly spin a conductive sphere\cite{schuck_ultrafast_2018}. Also spinning dipoles can be used for contactless manipulation of the position and orientation of metal spheres\cite{pham_dexterous_2021,tabor_adaptive_2022,dalton_attracting_2022}.
%This suggests eddy currents could be key.

A number of papers analytically derived the eddy currents and associated torques in special cases, for conducting spheres spinning in constant B-fields\cite{hertz_induction_1880,reichert_complete_2012,mcdonald_conducting_nodate,youngquist_slowly_2016,nurge_thick-walled_2017,nurge_drag_2018}. In particular, building on the work of Hertz\cite{hertz_induction_1880}, Nurge, Youngquist and co-workers\cite{youngquist_slowly_2016,nurge_thick-walled_2017,nurge_drag_2018} derived general solutions for all combinations of a solid or hollow sphere, uniform or azimuthally symmetric B-field and with the symmetry axis of $\mathbf{B}$ parallel or perpendicular to $\boldsymbol{\omega}$. That is, the B-field has to be rotation symmetric around an axis through the spheres center. This would be the case if $\mathbf{m}_\text{r}$ was parallel the displacement vector between the magnets, as studied by Yu et.\ al.\cite{yu_electromagnetic_2021,yu_optimal_2022}, but not in the present geometry. Nonetheless, we can show analytically that the frequency locking mechanism is present.
Here it should be noted that Reichert et.\ al.\cite{reichert_complete_2012} do consider a rotating B-field, but only by assuming that the constant field analysis applies in a coordinate system which co-rotates with the B-field. Below we prove explicitly that the relative frequency of field and spinning sphere is the key quantity.

We assume for this derivation that the rotor is directly above the floater ($\mathbf{d}=d\mathbf{\hat{z}}$), that $\mathbf{\hat{m}}_\text{r}$ is horizontal at all times ($\mathbf{\hat{m}}_\text{r} \mathbf{\cdot} \mathbf{\hat{z}} = 0$) and both magnets precess around vertical ($\boldsymbol{\hat{\omega}}_\text{r} = \boldsymbol{\hat{\omega}}_\text{f} = \mathbf{\hat{z}}$).
By Taylor expanding the rotor fields (see main text) in floater radius per levitation distance, $R_\text{f}/d$, we find
\begin{align}
   \mathbf{B}_\text{r} &= \frac{\mu_0}{4\pi r'^3} \left[3(\mathbf{\hat{r}'} \boldsymbol{\cdot} \mathbf{m}_\text{r}) \mathbf{\hat{r}'} - \mathbf{m}_\text{r}\right]  = \mathbf{B}_0 + \mathcal{O}\left(\frac{R_f}{d^4}\right)
   \notag\\
   \mathbf{E}_\text{r} &= \frac{\mu_0}{4\pi} \frac{\mathbf{\hat{r}'} \times \mathbf{\dot{m}_\text{r}}}{r'^2} = \mathbf{E}_0 + \mathbf{E}_1 + \mathcal{O}\left(\frac{R^2_f}{d^4}\right)   \label{eq:EM_fields_supp}
\end{align}
where
\begin{align*}
    \mathbf{B}_0 = \mathbf{B}_\text{r}(\mathbf{r} = 0) = -\frac{\mu_0}{4\pi d^3} \mathbf{m}_\text{r} \quad \text{,} \quad \mathbf{E}_0 = \frac{\mu_0}{4\pi d^3} \mathbf{\dot{m}}_\text{r} \times \mathbf{d}
\end{align*}
are uniform, $B_0 = |\mathbf{B}_0|$, and
\begin{align*}
    \mathbf{E}_1 = (3\mathbf{z} - \mathbf{r}) \times \mathbf{\dot{B}}_0.
\end{align*}
% We can divide into 2 problems : spinning, conductive sphere in B-field and conductive sphere in spinning E-field.
% For $\mathbf{B}_0$ we can apply the exact analysis of Ref. \cite{nurge_thick-walled_2017}, and
$\mathbf{E}_0$ induces an electric polarisation in the floater, but no eddy currents, so we focus on $\mathbf{B}_0$ and $\mathbf{E}_1$. This problem is equivalent to the floater spinning in a uniform magnetic field.
%For example, if the floater was centered on the rotation axis of a long, rotating solenoid the model would be nearly exact.
For example, if the floater was at the center of a long solenoid in the $xy$-plane which rotates around vertical, the model would be nearly exact.

% \textcolor{red}{Checkpoint. Næste del er meget midlertidig.}

% This is very close to the corresponding boundary value problem for $\mathbf{B}_0$ \cite[Eqs. 17-18]{nurge_thick-walled_2017}.

Since charges cannot leave the floater, we have $\mathbf{J} \mathbf{\cdot} \mathbf{r} = 0$ on the surface. In addition, we assume steady state, so that the charge distribution is constant, i.e.\ $\dot{\rho} = -\boldsymbol{\nabla} \mathbf{\cdot} \mathbf{J} = 0$. Inserting $\mathbf{E} = \mathbf{E}_1 - \boldsymbol{\nabla} V$ in Eq. \ref{eq:J_eddy} and neglecting terms of order $R_\text{f}^2/d^4$, this yields the boundary value problem
% \begin{align}
%     &\nabla^2 V = \boldsymbol{\nabla} \mathbf{\cdot} (\mathbf{E}_1 + \mathbf{v} \times \mathbf{B}_0) = 0
%     \notag\\
%     &\partial_r V_{r=R_\text{f}} = 3 \mathbf{\hat{r}} \mathbf{\cdot} (\mathbf{z} \times \mathbf{\dot{B}}_0) - B_0 \omega_\text{f}(\mathbf{\hat{m}}_\text{r} \mathbf{\cdot} \mathbf{r}) \mathbf{\hat{z}}_{r=R_f} \label{eq:Boundary_value_problem}
% \end{align}
\begin{align}
    \nabla^2 V = \boldsymbol{\nabla} \mathbf{\cdot} (\mathbf{E}_1 + \mathbf{v} \times \mathbf{B}_0) = 0
    \quad \text{and} \quad
    \partial_r V\vert_{r=R_\text{f}} = \mathbf{\hat{r}} \mathbf{\cdot} (\mathbf{E}_1 + \mathbf{v} \times \mathbf{B}_0) \vert_{r=R_\text{f}} \label{eq:Boundary_value_problem}
\end{align}
where the boundary condition refers to the fields and potential inside the floater, while the external potential is found by continuity of $V$. This is a Laplace equation with a Neumann boundary condition.

Without loss of generality, we pick coordinates such that $\mathbf{\hat{m}}_\text{r} = \mathbf{\hat{x}}$, in which case $\boldsymbol{B}_0 = - B_0 \mathbf{\hat{x}}$. It follows that
\begin{align}
    \mathbf{E}_1 &= (3\mathbf{z} - \mathbf{r}) \times (\boldsymbol{\omega}_\text{r} \times \mathbf{B}_0) = -\omega_\text{r} B_0 (2 \mathbf{z} - \mathbf{x} - \mathbf{y}) \times (\mathbf{\hat{z}} \times \mathbf{\hat{x}}) = \omega_\text{r} B_0 [2z \mathbf{\hat{x}} + x \mathbf{\hat{z}}] \label{eq:E_1}
    \\
    \mathbf{v} \times \mathbf{B}_0 &= [\boldsymbol{\omega}_\text{f} \times \mathbf{r}] \times \mathbf{B}_0 = -\boldsymbol{\omega}_\text{f} (\mathbf{B}_0 \cdot \mathbf{r}) + \mathbf{r} (\mathbf{B}_0 \mathbf{\cdot} \boldsymbol{\omega}_\text{f}) = \omega_\text{f}B_0 x \mathbf{\hat{z}}  \label{eq:v_cross_B}
\end{align}
We can thus write the boundary condition in spherical coordinates as
\begin{align*}
    \partial_r V\vert_{r=R_\text{f}} = (3 \omega_\text{r} + \omega_\text{f})B_0 R_\text{f} \sin\theta \cos\theta \cos\phi
\end{align*}
Interestingly, this is identical to \cite[Eq.\ 18]{nurge_thick-walled_2017} except for a different prefactor, even though we include an induced $\mathbf{E}$-field and use a different coordinate system.
The solution is
\begin{align*}
    V(r, \theta, \phi) = \frac{3 \omega_\text{r} + \omega_\text{f}}{2} B_0 \sin\theta \cos\theta \cos\phi \cdot
    \begin{cases}
        r^2 & r \leq R_\text{f}
        \\
         r^{-3} R^5_\text{f} & r \geq R_\text{f}
    \end{cases}
\end{align*}
as may be verified by inserting in Eq.\ \eqref{eq:Boundary_value_problem}.
The resulting potential gradient can be expressed in Cartesian coordinates as
\begin{align}
    \boldsymbol{\nabla} V\vert_{r < R_\text{f}} &= \frac{3\omega_\text{r} + \omega_\text{f}}{2} B_0 r(2 \sin \theta \cos \theta \cos \phi \mathbf{\hat{r}} + \cos 2\theta \cos \phi \boldsymbol{\hat{\theta}} - \cos \theta \sin \phi \boldsymbol{\hat{\phi}})
    \notag \\
    &= \frac{3\omega_\text{r} + \omega_\text{f}}{2} B_0 (z \mathbf{\hat{x}} + x\mathbf{\hat{z}})   \label{eq:grad_V}
\end{align}
The surface charge is given by
\begin{align}
    \partial \rho &= \epsilon_0\left[-\boldsymbol{\nabla} V(r > R_\text{f})\vert_{r = R_\text{f}} + \boldsymbol{\nabla} V(r < R_\text{f})\vert_{r = R_\text{f}}\right]
    \notag\\
    &= \frac{5}{2}\epsilon_0(3\omega_\text{r} + \omega_\text{f}) B_0 R_\text{f} \sin\theta \cos \theta \cos\phi  \label{eq:surface_charge_density}
\end{align}
Now, using Eqs. \eqref{eq:E_1},\eqref{eq:v_cross_B} and \eqref{eq:grad_V}, the eddy current is found to be
\begin{align}
    \mathbf{J}_\text{eddy} &= \sigma (\mathbf{E}_1 - \boldsymbol{\nabla} V + \mathbf{v} \times \mathbf{B}_0) = \frac{1}{2}\sigma(\omega_\text{r} - \omega_\text{f})B_0[z \mathbf{\hat{x}} - x \mathbf{\hat{z}}]
    \notag\\
    &= \frac{1}{2}\sigma(\omega_\text{r} - \omega_\text{f})[(\mathbf{B}_0 \boldsymbol{\cdot} \mathbf{r})\mathbf{\hat{z}} - z \mathbf{B}_0]    \label{eq:J_eddy_solved}
\end{align}
The physical interpretation of the preceding calculations is that charges accumulate on the floater surface, yielding a potential gradient which redirects the internal current, and this prevents further charge accumulation.

The current in Eq.\ \ref{eq:J_eddy_solved} induces its own B-field, which induces its own eddy current etc.\ ad infinitum. But we observe that Eq. \ref{eq:J_eddy_solved} is identical to \cite[Eq. 21]{nurge_thick-walled_2017} except $\omega \xrightarrow{} \omega_\text{f} - \omega_\text{r}$, so we could in principle read off the self-consistent solution in Ref.\ \cite{nurge_thick-walled_2017}. That said, Ref. \cite{nurge_thick-walled_2017} finds that the self-induction is negligible when
%$R_\text{f} \lesssim 2 d_\text{skin}$ where
\begin{align*}
    R_\text{f} \lesssim 2 d_\text{skin} \quad \text{where} \quad d_\text{skin} = (\mu_0 \sigma |\omega_\text{f} - \omega_\text{r}|)^{-1/2}.
\end{align*}
We note that $d_\text{skin}$ is the low-frequency limit of the electromagnetic skin depth\cite{griffiths_introduction_2013}. In our experiments $d_\text{skin} \geq 30\: \mathrm{mm}$ which is larger than our typical floater sphere which has a radius of 13 mm.
%$R_\text{f} = 13\: \mathrm{mm}$.
We conclude that the eddy-currents are bulk currents, self-induction is negligible and we can get the torque directly from Eq. \ref{eq:J_eddy_solved} :
\begin{align}
    \boldsymbol{\tau}_\text{eddy} &= \int_{\text{floater}} \mathbf{r} \times (\mathbf{J}_\text{eddy} \times \mathbf{B}_0) \mathrm{d} \mathbf{r} = \frac{2\pi}{15} \sigma (\omega_\text{r} - \omega_\text{f}) B^2_0 R^5_\text{f} \mathbf{\hat{z}} \label{eq:tau_eddy}
\end{align}
The net force on eddy currents is 0.

Eq. \ref{eq:tau_eddy} has two terms. One is a pure damping torque proportional to $\omega_\text{f}$ which results from the conductive floater spinning in a B-field. This damping torque is well-studied\cite{nurge_thick-walled_2017,lorrain_magnetic_1998,hertz_induction_1880}. The second is a purely accelerating torque proportional to $\omega_\text{r}$, which results from the E-field induced by the spinning rotor, or more precisely $\mathbf{B}_\text{0}$ pulling on the eddy currents induced by the E-field. To our knowledge, the accelerating torque has not been calculated previously.

From Eq. \eqref{eq:J_eddy_solved}, it is seen that when $\omega_\text{f} = \omega_\text{f}$, then $\mathbf{J}_\text{eddy} = 0$, i.e.\ when rotor and floater rotate at the same rate around the same axis, the two sources of eddy current precisely cancel. This is consistent with the result due to Backus that when a given volume of magnetic material rotates like a rigid body, no eddy currents are induced\cite{backus_external_1956}.

In deriving Eq. \ref{eq:tau_eddy} we have neglected translation, assumed that the rotor and floater both rotate around an axis through their centers and that the rotor moment is perpendicular to said axis. In general, there may be a B-field component parallel to $\boldsymbol{\hat{\omega}}_\text{f}$, however this does not produce an eddy current torque\cite{nurge_thick-walled_2017}. Also the accelerating and decelerating torques are not in general parallel, resulting in more complicated dynamics than frequency locking. Finally we neglected higher order corrections from the non-uniformity of $\mathbf{B}_\text{r}$. That said, Eqs. \eqref{eq:J_eddy_solved} and \eqref{eq:tau_eddy} is a leading order model, so it can be used to estimate the order-of-magnitude, time-scale and parameter scaling of eddy current effects, as we do in the main text.

\section{Electrodynamic torques \label{subsec:Electrodynamic_torques}}

The eddy current torque from Eq.\ \eqref{eq:tau_eddy} is discussed in the main text. Here we show that all other torques resulting from electrodynamic effects are entirely negligible. We use $\omega$, $m$ and $R$ for characteristic angular velocity, magnetic moment and radius of both floater and rotor, and $B$ for the characteristic size of the rotor B-field. In our experiments, a high frequency is $\omega \sim 1000 \: \mathrm{Hz}$, a typical radius is $R \sim 10\: \mathrm{mm}$ and the dipole field from rotor on floater when the magnets touch is around $B \sim 100\:\mathrm{mT}$.

In \ref{subsec:eddy_current_coupling} we found that eddy currents lead to a surface charge density which scales as $\partial \rho \sim \epsilon_0 R \omega B$. There is a corresponding convective surface current $\mathbf{K}_\text{conv} = \partial \rho \mathbf{v} \sim \epsilon_0 R^2 \omega^2 B$ from the charges moving along with the mechanical rotation of the floater. This leads to the torque $\boldsymbol{\tau}_\text{conv} = \int_\text{surf} \mathbf{r} \times (\mathbf{K}_\text{conv} \times \mathbf{B}) \mathrm{d}^2 \mathbf{r} \sim \epsilon_0 \omega^2 R^5 B^2$. Comparing to the dipole torque, $\boldsymbol{\tau}_\text{dip} = \mathbf{m} \times \mathbf{B} \sim B^2R^3/\mu_0$, we find
\begin{align*}
    \frac{\tau_\text{conv}}{\tau_\text{dip}} \sim \mu_0 \epsilon_0 R^2 \omega^2 = \left(\frac{R\omega}{c}\right)^2 \approx 10^{-15}
\end{align*}
For a conducting sphere spinning parallel to a uniform B-field, there is instead a charge density of the form\cite{nurge_thick-walled_2017} $\rho = -2\epsilon_0 \omega B$, but the torque has the same approximate magnitude.

Another torque is from the induced E-field acting on the charges. The dipolar E-field scales as $E_\text{dip} \sim \mu_0 m \omega/d^2 \sim R^3\omega B /d^2$. Thus $\tau_\text{$E$-on-$\rho$} = \int \rho \mathbf{r} \times \mathbf{E} \mathrm{d} \mathbf{r} \sim \epsilon_0 \omega^2 R^5 B^2 \frac{R^2}{d^2}$, which is even smaller than $\boldsymbol{\tau}_\text{conv}$. The same conclusion is found for $\mathbf{E}_\text{dip}$ acting on surface charges. Similar relations may be found for the forces on induced charges and convective currents, and the arguments hold equally whether the charges and currents are induced in the floater or the rotor. We conclude, in agreement with the literature\cite{lorrain_electrostatic_1990,lorrain_magnetic_1998,mcdonald_conducting_nodate,youngquist_slowly_2016}, that convective currents and forces on induced charges are entirely negligible when the floater surface moves at non-relativistic speeds.

Finally, the dipole electric fields induce an electric polarisation in the magnets. For a conductive sphere in a uniform electric field, the surface charge distribution is\cite{griffiths_introduction_2013} $\partial \rho = 3 \epsilon_0 E \cos \theta$. Using the zeroth order field from Eq. \eqref{eq:EM_fields_supp} we find that $\partial \rho_{E_0} \sim \epsilon_0 E_0 \sim \epsilon_0 d \omega B_0 \sim \frac{d}{R_\text{f}} \partial \rho$. That is, the charge density induced by $E_0$ is comparable to that from the eddy current problem, so by the same argument all resulting forces and torques are negligible when $\omega R \ll c$.

\section{Additional simulations}

\end{strip}
\renewcommand{\thefigure}{3}
\renewcommand{\figurename}{Figure Supp.}
\begin{figure*}[!h]
    \subfigure{\includegraphics[width=0.49\textwidth]{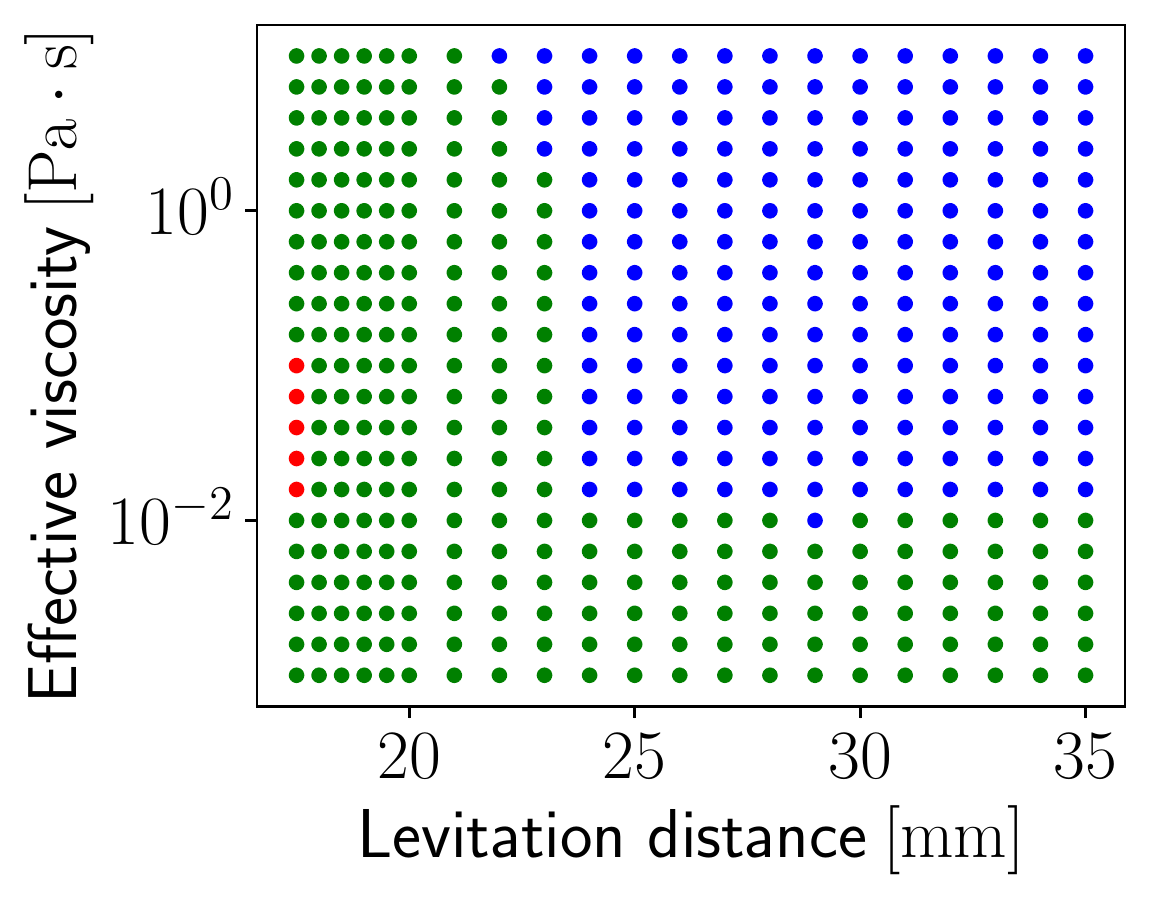}}
    \subfigure{\includegraphics[width=0.49\textwidth]{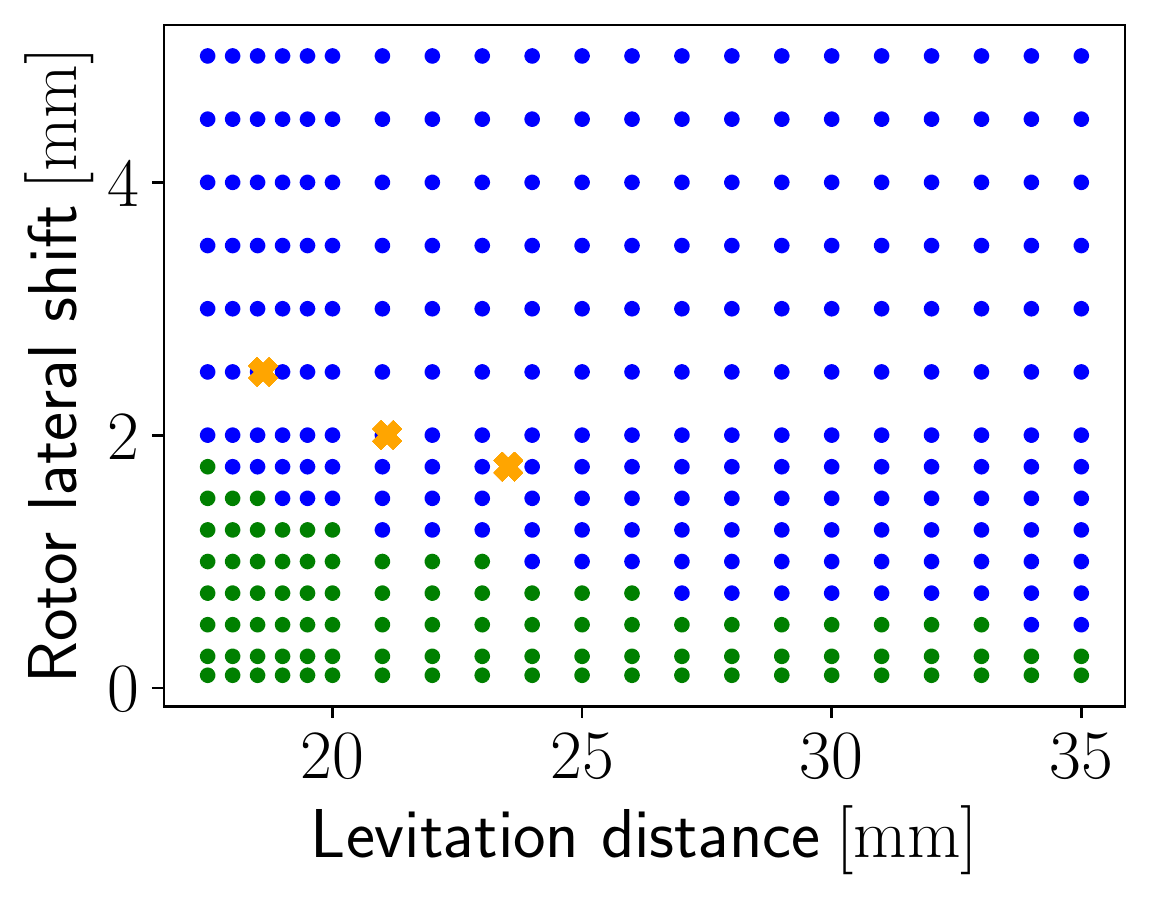}}

    \subfigure{\includegraphics[width=0.49\textwidth]{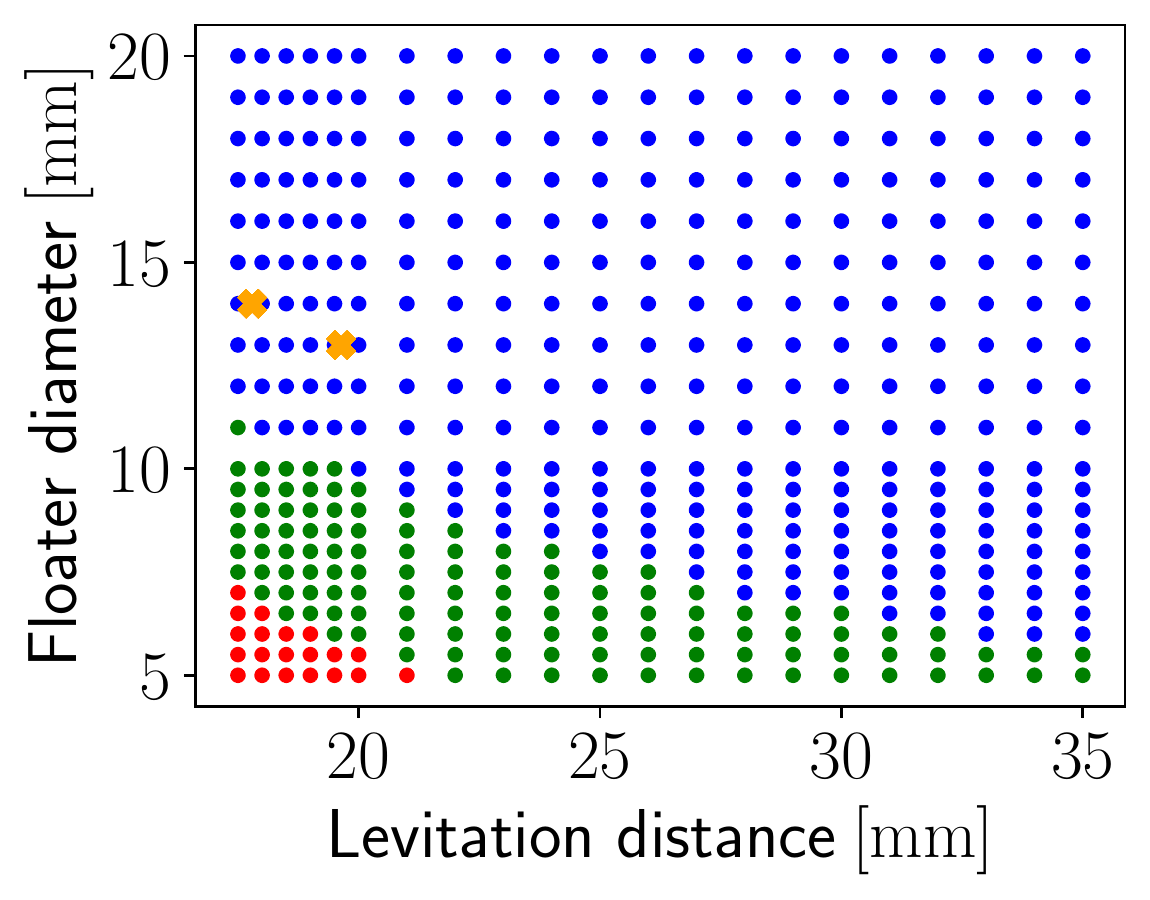}}
    \subfigure{\includegraphics[width=0.49\textwidth]{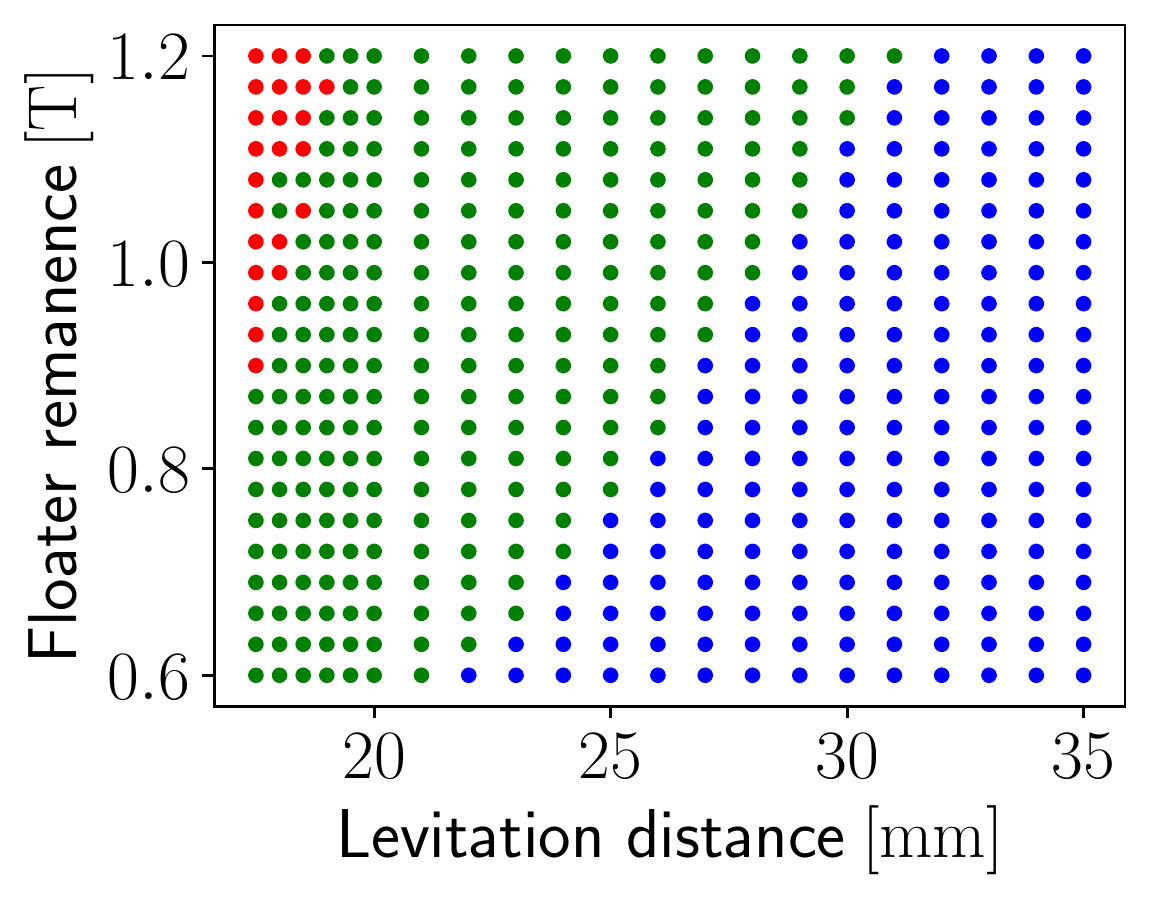}}
    \caption{Distribution of the 3 dynamical phases in parameter space. Characterised by steady-state behaviour for simulations at fixed levitation distance $d$. Orange crosses are predicted points of stability. See main text for details. Unless varied, the default parameters are remanences of $M_\text{f} = M_\text{r} = 1.18 \: \mathrm{T}$, viscosity of $\eta = 1\: \mathrm{Pa \cdot s}$, floater diameter of 12.7 mm, rotor magnet diameter of 19.7 mm and rotor lateral shift of $\delta_\text{r} = 1 \: \mathrm{mm}$. The rotor frequency is $f_\text{r} = 200 \: \mathrm{Hz}$ for the viscosity- and lateral shift variations, 300 Hz for floater diameter and 150 Hz for remanence.}
    \label{Fig.dynamical_phases_supp}
\end{figure*}

\begin{strip}

Fig. \ref{Fig.dynamical_phases_supp} shows the simulated, steady-state behaviour of the floater magnet for different parameters. By levitation distance we mean the distance between rotor and floater along the rotation axis, but since $d \ll \delta_\text{r}$ this approximately equals the absolute distance. By effective viscosity we simply mean that it takes a far greater value than for air, which we attribute to eddy current damping. In the main text we present and discuss the variation with rotor frequency (Fig. 7). The frequencies for the diameter and remanence variation are chosen to match the experiments (Fig. 6).

As explained in the main text, we understand the phases as follows :
Red is the energy minimum, i.e.\ where rotor and floater moments are always anti-parallel, hence frequency locked. In the green phase, the floater cannot keep up with the rotor, which results in 2 precessional modes. A slow, large-amplitude mode and a small-amplitude one at the rotor frequency. As the frequency increases further, the constant, vertical magnetic field becomes more important \textit{relative} to the rotating, horizontal field. Therefore as one increases rotor frequency, the floater moment becomes increasingly vertical. At some critical point, the two precessional modes recombine. Then the polar angle is constant, and frequency-locking is reestablished with the floater moments horizontal component nearly parallel to the rotor moment. This is the blue phase. We understand constancy of the polar angle as a result of gyroscopic stability, which relies on the moment of inertia.

This explanation is consistent with all the data in Fig \ref{Fig.dynamical_phases_supp}. Increasing the lateral shift increases the constant field, hence making the blue phase more prevalent. Decreasing levitation distance or increasing remanence increases the dipole torque, which facilitates the floater keeping up with the rotor, thus the blue phase becomes less prevalent. Drag makes the floater lag behind the rotor, reducing the susceptibility to the rotating field, so unsurprisingly increased viscosity facilitates the blue phase. Finally, there is floater diameter, which increases rotational drag as $R_\text{f}^3$, moment of inertia as $R_\text{f}^5$ and floater moment hence dipole coupling strength as $R_\text{f}^3$. We observe that a larger floater is more prone to change phase from red to green to blue, so the increase in drag and inertia win out over the increased dipole coupling; perhaps because of the moment of inertia's stronger size dependence.

\section{Fixed floater position : further analysis}

Here we derive analytical expressions for the steady-state moment orientation and angular velocity in the blue phase, i.e.\ the one we find to enable levitation. We assume the floater position is fixed vertically below the rotor.

Let $\mathbf{\hat{m}}_\text{r}(t = 0) = \mathbf{\hat{x}}$. The rotor field and floater moment may then be written
\begin{align*}
    \mathbf{B}_\text{r} =
    \begin{pmatrix}
    -B_{\text{r}, \perp} \cos(\omega_\text{r} t) \\ - B_{\text{r}, \perp}\sin(\omega_\text{r} t) \\ B_z
    \end{pmatrix}
    \quad \text{and} \quad
    \mathbf{m}_\text{f} = m_f
    \begin{pmatrix}
        \sin \theta_\text{f} \cos(\omega_\text{r} t + \varphi) \\ -\sin \theta_\text{f} \sin(\omega_\text{r} t + \varphi) \\ \cos \theta_\text{f}
    \end{pmatrix},
\end{align*}
where $B_{\text{r}, \perp}$ is the rotating field component in the $xy$-plane and $B_{\text{r}, z}$ is the constant, vertical field resulting from imperfections in rotor placement. The remaining variables are defined in Fig. 2 of the main text. Note the clockwise rotation.

As pointed out in the main text, the blue phase is characterised by $\dot{\varphi} = \dot{\theta}_\text{f} = 0$, so the angular velocity must have the general form $\boldsymbol{\omega}_\text{f} = -\omega_\text{r} \mathbf{\hat{z}} + \dot{\psi} \mathbf{\hat{m}}_\text{f}$. The first term is precession around vertical, the second spinning around the magnetic moment. Any other angular velocity contribution would cause a time-variation in either $\varphi$ or $\theta_\text{f}$. Additionally we have the relation $\mathbf{\hat{m}}_\text{f} \cdot \boldsymbol{\omega}_\text{f} = 0$ which holds for all our simulations when the floater starts from rest. It follows that $\dot{\psi} = \omega_\text{r} \cos \theta_\text{f}$, so
\begin{align*}
    \boldsymbol{\omega}_\text{f} = \omega_\text{r} \left[-\mathbf{\hat{z}} + \cos \theta_\text{f} \mathbf{\hat{m}}\right] = \frac{\omega_\text{r}}{2}
    \begin{pmatrix}
    \sin(2\theta_\text{f}) \cos(\omega_\text{r} t + \varphi) \\ -\sin(2\theta_\text{f}) \sin(\omega_\text{r} t + \varphi) \\ -2 \sin^2 \theta_\text{f}
    \end{pmatrix},
\end{align*}
and the angular acceleration is
\begin{align*}
    \boldsymbol{\dot{\omega}}_\text{f} = -\frac{\omega_\text{r}^2}{2} \begin{pmatrix}
    \sin(2\theta_\text{f}) \sin(\omega_\text{r} t + \varphi) \\ \sin(2\theta_\text{f}) \cos(\omega_\text{r} t + \varphi) \\ 0
    \end{pmatrix}.
\end{align*}
The two equations above are analogous to Ref. \cite[Eqs. 17-18]{ucar_polarity_2021}, however we disagree on the exact form.

Now consider the governing equation:
\begin{align*}
    I_\text{f} \boldsymbol{\dot{\omega}}_\text{f} = \mathbf{m}_\text{f} \times \mathbf{B}_\text{r} - \zeta_\text{rot} \boldsymbol{\omega}_\text{f}.
\end{align*}
Without loss of generality, we insert all of the above vector formulas at $t=0$, which gives the 3 coupled equations
\begin{align*}
    -\frac{I_\text{f} \omega_\text{r}^2}{2}
    \begin{pmatrix}
    \sin(2\theta_\text{f}) \sin \varphi \\ \sin(2\theta_\text{f}) \cos \varphi \\ 0
    \end{pmatrix}
    = -m_\text{f}
    \begin{pmatrix}
        B_{\text{r}, z} \sin \theta_\text{f} \sin \varphi \\ B_{\text{r}, z} \sin \theta_\text{f} \cos \varphi + B_{\text{r}, \perp} \cos \theta_\text{f} \\ B_{\text{r}, \perp} \sin \theta_\text{f} \sin \varphi
    \end{pmatrix}
    +\frac{\zeta_\text{rot} \omega_\text{r}}{2}
    \begin{pmatrix}
        -\sin 2\theta_\text{f} \cos \varphi \\ \sin 2\theta_\text{f} \sin \varphi \\ 2 \sin^2 \theta_\text{f}
    \end{pmatrix}.
\end{align*}
The equations can in principle be solved numerically for $\theta_\text{f}$ and $\varphi$, but instead we use that $\theta_\text{f}, \varphi$ are very small in the blue phase to solve analytically. To leading order we have
\begin{align*}
    I_\text{f} \omega_\text{r}^2
    \begin{pmatrix}
    \varphi \\ \theta_\text{f} \\ 0
    \end{pmatrix}
    = m_\text{f}
    \begin{pmatrix}
        B_{\text{r}, z} \varphi \\ B_{\text{r}, z} \theta_\text{f} + B_{\text{r}, \perp} \\ B_{\text{r}, \perp} \varphi
    \end{pmatrix}
    -\zeta_\text{rot} \omega_\text{r}
    \begin{pmatrix}
        -1 \\ \theta_\text{f} \varphi \\ \theta_\text{f}
    \end{pmatrix}.
\end{align*}
The solution is
\begin{align}
    \varphi = \frac{\zeta_\text{rot} \omega_\text{r}}{I_\text{f}\omega_\text{r}^2 - mB_{\text{r}, z}} \quad \text{and} \quad \theta_\text{f} = \frac{mB_{\text{r}, \perp}}{I_\text{f} \omega_\text{r}^2 - mB_{\text{r}, z}}.
\end{align}
We note that when $\theta_\text{f} \neq 0$, the general $z$-equation yields
\begin{align*}
    \sin \varphi = \frac{\zeta_\text{rot} \omega_\text{r}}{m_\text{f} B_{\text{r}, \perp}} \sin \theta_\text{f}
\end{align*}
which means that in the absence of damping, the steady-state value of $\varphi$ is exactly $0^{\circ}$ or $180^{\circ}$.

\newpage
%\bibliographystyle{unsrt}
%\bibliography{References}

\iffalse

\fi

\end{strip}

\end{document}